\documentclass[preprint,prd,preprintnumbers,nopacsnumbers,amsmath,amssymb]{revtex4-1}
\newcommand{\lbl}[1]{\label{#1}}

\usepackage{bm}% bold math

\usepackage{amsmath}
\usepackage{graphicx}
\usepackage{amssymb}
\usepackage{mathrsfs}
\usepackage{bbm}
\newcommand{\be}{\begin{eqnarray}}
\newcommand{\ee}{\end{eqnarray}}

\newcommand{\inter}{{\mathrm{int}}}
\newcommand{\WS}{{{\mathrm{WS}}}}

\newcommand{\cL}{{\cal L}}

\newcommand{\bbbone}{{{\mathchoice {\rm 1\mskip-4mu l} {\rm 1\mskip-4mu l}
{\rm 1\mskip-4.5mu l} {\rm 1\mskip-5mu l}}}}

\newcommand{\vct}[1]{\vec{#1}}

\newcommand{\vecG}{\vct{G}}

\begin{document}

%
%%%%%%%%%%%%%%%%%%%%%%%%%%%%%%%%%%%%%%%%%%%%%%%%%%%%%%%%%%
%
%\pagenumbering{empty}
%\begin{titlepage}
%
\title{ On Relations Between Electroweak  Hierarchy Problem, Fluctuating Three  Branes 
And General Covariance}
  
%\vskip 5.0cm
\author{Antti J. Niemi}
\email{Antti.Niemi@physics.uu.se}
\affiliation{Department of Physics and Astronomy, Uppsala University,
P.O. Box 803, S-75108, Uppsala, Sweden}
\affiliation{
Laboratoire de Mathematiques et Physique Theorique
CNRS UMR 6083, F\'ed\'eration Denis Poisson, Universit\'e de Tours,
Parc de Grandmont, F37200, Tours, France}
\author{Sergey Slizovskiy} 
\email{Sergey.Slizovskiy@physics.uu.se}
\affiliation{Department of Physics and Astronomy, Uppsala University,
P.O. Box 803, S-75108, Uppsala, Sweden}

\begin{abstract}
\noindent
We inquire whether a resolution to the electroweak hierarchy problem could reside in 
symmetries that relate the bosonic Weinberg-Salam Lagrangian with a
higher dimensional generally covariant theory. 
For this we consider a three-brane that moves under the influence 
of a seven dimensional pure Hilbert Einstein-like generally covariant theory.  We introduce  a change of variables that combines 
the conformal scale of the metric tensor with the brane fluctuations,
so that the conformal scale becomes the modulus and the fluctuations
become the angular field degrees of freedom of a polarly decomposed  Higgs. 
When we  assume that the  four dimensional  space-time background of the generally covariant
theory is locally conformally flat and that the internal space is
a squashed three-sphere, we arrive at one massless and 
three massive vector fields akin those in the Weinberg-Salam model  
and recover all the familiar ingredients of the symmetry broken 
bosonic Weinberg-Salam model,  except that there is no bare Higgs mass. The Higgs mass
is subject to dimensional censorship, its presence  is forbidden  by  general covariance.  
This proposes that a resolution to the  electroweak hierarchy problem 
might  well reside in higher dimensions, in a Ward-Takahashi like identities for
general covariance that relate  a non-vanishing Higgs mass to  dynamical breaking of general covariance.
Moreover,  the two electroweak gauge couplings are both determined by the 
squashing parameter of the internal three sphere and
when we impose the condition  that the vector boson masses must be in line with  custodial symmetry 
we arrive at the classical level to the Weinberg angle $\sin^2 \! \theta_W \approx  0.296 $.
\end{abstract}

\maketitle

\section{Introduction}
\vskip 0.5cm

Experiments at  LHC should soon expose the Higgs particle and reveal the mechanism of the
electroweak symmetry breaking. Eventually LHC should also allow us to scrutinize 
the fine details of the Higgs sector and give some hints
how Physics behaves  at even higher energy scales.   A {\it Big Issue}  is the inherent instability of the 
mass of the standard electroweak Higgs.  In the Weinberg-Salam model the natural value of the Higgs  
mass  is proportional to the cut-off  scale of the theory  \cite{wel09}, \cite{pol92},  thus a delicate fine-tuning  
becomes necessary in order for it to attain realistic values  \cite{giu08}. As a consequence the predominant point of view is 
that the Weinberg-Salam model is an effective low energy theory, valid  only to a scale no higher than a few
$TeV$. Several approaches are pursued to resolve this  electroweak hierarchy problem. These include
supersymmetric  theories \cite{dine}, little Higgs models \cite{little}, extra dimensions \cite{sun} and many others. 

In the present article we shall explore (perhaps) new ways to address the electroweak  hierarchy problem. We search for a resolution
from such symmetries of the  Weinberg-Salam Lagrangian that have a higher dimensional origin and interpretation. 
Our modus operandi is the Kaluza-Klein approach  \cite{pope}, \cite{appel}. 
But instead of attempting to formulate the electroweak theory in a higher dimensional space-time as in a
conventional Kaluza-Klein approach, we restrict our attention to symmetry
structures in the four dimensional electroweak theory that allow for a higher dimensional Kaluza-Klein interpretation: 
The conventional Kaluza-Klein approach leads to 
still unresolved problems in particular in 
the fermionic sector \cite{witten},  and for this reason we prefer not to start from any {\it a priori}  prescribed higher dimensional theory.  
In our view, the question whether we live in a  higher dimensional space-time  or not  remains  a rather interpretational and maybe even philosophical
one.  A resolution may well be in a hybrid picture where bosonic variables fluctuate
into higher dimensions while fermions are constrained into four, if indeed a higher dimensional interpretation that
goes beyond pure symmetry considerations turns out to have some real advantages.  

Our point of view is rather pragmatic and of diagnostic nature: We  inquire what  is a minimal amount of higher
dimensional structure present in the Weinberg-Salam model that we need to look into, in order to properly address and 
hopefully resolve the electroweak hierarchy problem in the limited context of 
the four dimensional bosonic Weinberg-Salam Lagrangian.  
For this we  consider a mathematical construct, that combines a four-dimen\-sional space-time 
manifold $\mathbb M_4$ with some hypothetical higher dimensional  internal manifold. It appears likely  that whatever the
structure of the internal manifold, it should somehow relate to the  squashed three-sphere    
since on a squashed three-sphere the commutators of Killing vectors coincide  with the Lie algebra of $SU(2) \times U(1)$  \cite{pope}. 
Consequently  we think it is reasonable to assume 
that {\it any} approach to electroweak interactions that involves higher dimensions,  somehow  relates to the structure of 
a squashed three sphere, and thus we take it to be our internal manifold.

In our  higher dimensional construct, we then   bring in a generally covariant theory with an action similar to that of 
pure Einstein gravity (even though we make no claims that the theory has anything to do with gravity -  our focus is on
general covariance and the "gravity" is simply a mathematical construct to realize it). 
We inspect its symmetry structure in a background that is
determined by  a  suitably chosen metric tensor that we subject to a  Kaluza-Klein dimensional reduction onto 
massless modes. Since we are only interested in how symmetries with a higher dimensional interpretation act on the field content of the
Weinberg-Salam model,  we can safely ignore any issues related to complications that may arise due to dilaton fields 
or from  an infinite tower of massive modes.    Questions on higher dimensional
stability of the construction are also too technical and complex to be addressed here, 
for the present purposes it suffices  to note that  experimental observations show no sign of instability in the standard
electroweak theory.

From the  conventional point of view of a Kaluza-Klein compactification our approach can be interpreted so that we are inquiring
how to {\it minimally embed } the Weinberg-Salam Lagrangian into a higher dimensional context,  rather than {\it deriving} it from
a consistent Kaluza-Klein truncation and with no additional fields.  We find it quite 
unlikely that a full and consistent truncation with no added four dimensional fields  is even possible, at least we are not familiar with any. 
Instead of addressing this issue which is at the hearth of  any fully consistent 
higher dimensional approach,  we  take a minimalistic point of view  to simply query that whatever
the higher dimensional theory (if it indeed exists, also for fermions) it must lead to certain symmetry structures that we wish to expose and apply to address
the electroweak hierarchy problem. In this manner we arrive at a four dimensional $SU(2) \times U(1)$ gauge theory, which is essentially 
the standard electroweak theory: We recover the correct mass assignments for the vector fields,
but there is {\it no} Higgs mass  and  the two gauge couplings are  not independent but related and determined by the parameter 
describing  the squashing of the three-sphere. The reason why the  Higgs mass is absent is rather telling:
A bare Higgs mass term breaks general covariance.  Consequently, {\it if} there is a good reason to insist on general covariance 
as it appears here, there should be some Ward-Takahashi type identity that ensures that also in the quantized electroweak 
theory there is no Higgs mass term,   thus no hierarchy problem.  However, there is a  non-vanishing Higgs condensate that  
has a  dynamical origin in a $A^2$-condensation of the intermediate vector bosons \cite{guba1}-\cite{fad2}. 

It is notable that even though our construction leaves no room even for a primordial Higgs field,  we do obtain the correct masses for the 
$SU(2)\times U(1)$ gauge fields.  However,  the mechanism that equips the three intermediate bosons with their mass  
differs from the standard approach based on spontaneous breaking of $SU_L(2) \times U_Y(1)$ into electromagnetic $U(1)$ by the Higgs field:
We introduce a three-brane that in our mathematical construct asymptotically coincides with
the physical four dimensional space-time, but  can locally fluctuate  into the higher 
dimensional manifold where it moves under the influence of the higher dimensional 
Hilbert- Einstein like generally covariant interaction. The brane fluctuations become   "eaten up"
by the longitudinal modes of three vector bosons and as a consequence they become massive,  
their masses being determined by the brane tension in combination of the squashing parameter. 
We show that the brane fluctuations are precisely the angular components  of the Standard
Model Higgs field,  in a polar decomposition of the Higgs. The remaining field degree of freedom,   the Higgs modulus, resides in the
conformal scale of the four dimensional Hilbert-Einstein like generally covariant  Lagrangian. If we specify to a conformally flat space-time
we obtain by a change of variables a Lagrangian that is very much like  the original  Weinberg-Salam model in the conformally invariant
(Coleman-Weinberg) limit of the Higgs potential,  in the {\it flat}  space-time $\mathbb R^4$ and with the correct mass relations for the 
intermediate vector bosons.  Except that we now have also a Ward-Takahashi like condition that the underlying general 
covariance should remain unbroken, preventing the presence of a bare Higgs mass, also when radiative corrections 
are accounted for.

Furthermore, from the value $\alpha = 1/137$ of the fine structure constant
we estimate at the classical level and in the purely bosonic theory the  value 
\[
\sin^2 \theta_W \approx  0.296
\]
for the Weinberg angle.  This  is an experimental  constraint for our higher dimensional interpretation, and even though we derive
it in the  purely bosonic theory and at the classical level, the result   
is surprisingly close to the experimentally measured value  \cite{E158}, \cite{pdg}. Indeed,
the final theory is almost {\it verbatim} equal to the Weinberg-Salam model, except that the couplings now
have a common origin and there is no bare Higgs mass. 

\vskip 0.3cm

In the next section we  describe  our approach in the context of the Abelian Higgs Model, in  the wider context
of full Kaluza-Klein reduction.  Here there are no technical issues with the dilaton, and for completeness we 
display the entire dilaton sector.
We show how the massive $U(1)$ gauge boson and the neutral scalar that together determine 
the particle content of the Abelian Higgs model
in its  spontaneously broken phase,  emerge from the massless modes of a five dimensional Kaluza-Klein compactification 
{\it with no primordial Higgs field}. However,  the conventional  Higgs field  with the ensuing vector boson mass
can be fully reconstructed by introducing a three brane that locally
fluctuates into the fifth dimension with dynamics governed by the Nambu action.
The brane fluctuations are described by a variable that corresponds  to the phase of the Higgs field, and its 
modulus emerges from the conformal scale of the metric tensor when we consider the theory 
in a locally conformally flat space-time and introduce certain  changes of variables. After 
reconstructing the Higgs  we arrive at the standard classical Abelian
Higgs model in the space-time $\mathbb R^4$, with a Coleman-Weinberg type potential for the Higgs field at the classical level.
In the quantum theory there is a Higgs condensate it is determined by the $A^2$-condensation of the gauge field.
But since a bare Higgs mass breaks general covariance, a massive Higgs field  remains forbidden. 

Our full-fledged Kaluza-Klein analysis  that includes the dilaton sector reveals that while the dilatons have an essential
r\^ole in stabilizing the theory, the presence of dilatons is less important when we are only interested in relating the symmetry
structure of the higher dimensional theory with that of its Kaluza-Klein descendant. 
The dilaton has no effect on the emergence of vector mass in the dimensionally reduced theory, 
and in particular it has no effect on the Higgs mass that
is forbidden by general covariance. This gives us confidence that when we proceed to the non-Abelian case where we only
consider the symmetry structure, there is no reason to explicitely consider the effect of dilatons.

In Section 3 we describe the Weinberg-Salam model for the present purposes.  
In particular, we introduce a generalization of the changes of variables that in Section 2 enabled us to relate 
the Abelian Higgs model to a Kaluza-Klein reduction of five dimensional generally covariant theory with Hilbert-Einstein action.
In Section 4 we consider  certain mathematical properties of the squashed three sphere. We derive  a 
number of  relations that will be useful for us
when we proceed to construct the Weinberg-Salam model. 
In Section 5 we first show how to introduce the $SU(2) \times U(1)$ gauge structure of the Weinberg-Salam model,
by employing a Kaluza-Klein reduction that starts from a seven dimensional Hilbert-Einstein action  with  an  internal space that is a
squashed three sphere.  Following the Abelian example 
we explain how the vector fields acquire their masses when we introduce a three brane that is asymptotically
stretched into the non-compact directions, but is locally allowed to fluctuate
into the squashed three sphere.  We first consider a Nambu  action for the three brane. This yields us
a version of the Weinberg-Salam model with a wrong relation between the intermediate boson masses, 
and a  local $SU_L(2) \times SU_R(2)$ custodial symmetry which is explicitly broken. 
We then modify the Nambu action by introducing a parameter
that measures the deviation of the mass matrix from the point of custodial symmetry. When we choose this parameter 
to correspond to the value
where custodial symmetry is recovered 
we obtain the standard  Weinberg-Salam model in the Coleman-Weinberg limit of the Higgs potential but  {\it again}
with the additional Ward-Takahashi like condition  that a bare Higgs mass is forbidden by general covariance, and 
 the  couplings  corresponding to the gauge groups $SU(2)$ and $U(1)$ are 
determined by a single parameter  that describes the squashing of the three sphere. Furthermore,  we show that
from the familiar low energy value of the fine structure constant $\alpha = 1/137$ we find a value 
for the Weinberg angle that is quite close to the observed value. We also argue that  as in the Abelian case 
there is a  Higgs condensate that has its origin in the
$A^2$-condensation of the intermediate vector bosons.

Finally, in order to avoid the familiar  tricky issues  that are associated with the conformal scale and its  analytic continuation
\cite{gibb}  we work exclusively in a space-time with Euclidean
signature. However, we see no problems in extending our results to a space-time with Minkowskian metric: Since there is
no reason why the generally covariant Hilbert-Einstein  action that we introduce 
describes gravity, we may as well choose the overall sign of the Minkowskian scalar curvature to be opposite to that in Einstein gravity.

\section{Abelian Mass From Fluctuating Three Brane}
\lbl{vmfb}

\vskip 0.5cm

Here we show how the standard Abelian Higgs model
with its Higgs effect can be derived from the brane world, but without a primordial Higgs field.
We  start from 
the familiar, full  Kaluza-Klein decomposition of the five dimensional 
metric tensor \cite{pope}
\begin{equation}
ds^2 = g_{ij} dx^i dx^j = e^{2\alpha \phi} g_{\mu\nu} dx^\mu dx^\nu + 
e^{2\beta \phi} g_{55} \left(
d\theta + A_\mu dx^\mu\right)^2
\lbl{5dmetric}
\end{equation}
We label $i,j, ... = 0,...,4$ and $\mu,\nu, ... = 0,...,3$, and 
$\alpha$ and $\beta$ are parameters. 
The compact coordinate $\theta \in [0,2\pi)$ describes a circle
$\mathbb S^1$ with a local
radius that depends on the scalar field $\phi$ and the
constant $g_{55}$ has dimensions of length squared. 
Under a $x^\mu$ dependent reparametrization of $\mathbb S^1$ 
\begin{eqnarray}
\theta & \to & \theta + \varepsilon(x^\mu)
\\
\nonumber
A_\mu & \to & A_\mu - \partial_\mu \varepsilon
\lbl{repa}
\end{eqnarray}
the metric (\ref{5dmetric}) remains intact. For $A_\mu$ this is the familiar  $U(1)$ gauge transformation.

We are interested in the (low energy) limit 
where the fields become independent of the compact coordinate $\theta$.
In this limit we take the metric components $g_{\mu\nu}$ and the 
fields $\phi$, $A_\mu$ to depend only on the four dimensional coordinates
$x^\mu$. The ensuing five dimensional generally covariant
(Hilbert-Einstein) action with cosmological constant $\Lambda_{(5)}$
\[
S = \frac{1}{16\pi G_{(5)}}\int \! d^5x \sqrt{g_{(5)} } \, \left\{  R -  2 \Lambda_{(5)} \right\}
\]
becomes
\[
=\frac{1}{16\pi G} \int \! d^4x \sqrt{ g } \left[ e^{(\beta + 2\alpha) \phi} \left\{
R + 6\alpha(\alpha+\beta)  \nabla_\mu \phi\nabla^\mu\phi \right\} 
\right.
\]
\begin{equation}
\left. + \frac{1}{4}
e^{3\beta\phi} g_{55} F_{\mu\nu}F^{\mu\nu} - 2 \Lambda   e^{(\beta + 4\alpha)\phi} \right]
\lbl{HKK1}
\end{equation}
We have here removed a boundary term from partial integration,
$R$ is the Ricci scalar and $g$ is the determinant of the four dimensional $g_{\mu\nu}$, and $G$ is 
akin the four dimensional Newton's constant.

In the conventional approach \cite{pope}  one now proceeds by assuming that the parameters $\alpha$ and $\beta$
obey the following relation
\begin{equation}
\beta = - 2\alpha
\label{conv}
\end{equation}
But here  we proceed instead with  the complementary choice 
\[
\beta \not= -2\alpha
\] 
This allows us to introduce
the change of variables
\[
d\phi = \frac{1}{\beta + 2\alpha} \frac{d\sigma}{\sigma}
\]
which obviously can not be introduced if (\ref{conv}) is assumed. Next we define
\[ 
\kappa = \frac{\beta}{\beta + 2\alpha}
\]
and implement the conformal transformation 
\[
g_{\mu\nu} \to \sigma g_{\mu\nu}
\]
followed by the additional change of variables,
\[
d\chi = \sqrt{\frac{3}{2}} \kappa \frac{ d\sigma}{\sigma}
\]
In this way we find that (\ref{HKK1}) becomes 
\begin{equation}
= \frac{1}{16 \pi G} \int \! d^4x \sqrt{g } \left[
R - \nabla_\mu \chi \nabla^\mu\chi +  \frac{1}{4}
e^{\sqrt{6} \chi} g_{55} F_{\mu\nu}F^{\mu\nu} - 2 \Lambda  e^{-\sqrt{\frac{2}{3}} \chi} \right]
\lbl{HKK2}
\end{equation}
{\it independently} of $\alpha$ and $\beta$, provided of course that $\beta \not= -2\alpha$.
This  action has the same functional form as the four  dimensional Brans-Dicke action in interaction
with Maxwellian electrodynamics, with a coupling that depends on the dilaton field $\chi$. 
The Liouville-like instability of the dilaton ground state is apparent.

It is notable that since the dilaton has no charge the vector field remains massless.
In order to obtain a massive vector field and a relation to the Abelian Higgs model in its 
spontaneously broken phase 
we proceed to construct a gauge invariant mass term for the U(1) gauge field in (\ref{HKK2}). 
For this we consider a three-brane that stretches along the non-compact directions of the
five dimensional space-time. This brane is locally described by a scalar 
function 
\begin{equation}
\theta = h(x_\mu)
\lbl{htheta}
\end{equation}
The induced metric tensor on the brane is 
obtained by pulling-back the five-metric (\ref{5dmetric})
with the help of the basis vectors on the brane,
\begin{equation}
{E_\mu}^i = {\delta_\mu}^i + \frac{\partial h}{\partial x^\mu} {\delta_5}^i
\lbl{5bein}
\end{equation}
This gives the induced brane metric
\begin{equation}
G^{ind}_{\mu\nu} = {E_\mu}^i {E_\nu}^j g_{ij} = e^{2\alpha \phi} \left(
g_{\mu\nu} + g_{55} e^{2(\beta - \alpha) \phi }(A_\mu + \partial_\mu h) (A_\nu + \partial_\nu h)
\right)
\lbl{indbrane}
\end{equation}
Note that this metric is {\it invariant} under the $U(1)$ isometry (\ref{repa}): A local
shift in the brane position 
\[
h(x) \to h(x) + \varepsilon (x) 
\]
can be compensated by the shift 
\[
A_\mu \to A_\mu - \partial_\mu \varepsilon (x)
\]
in the gauge field.

We assume that the dynamics of the three-brane is governed by the Nambu action
\begin{equation}
S_{brane} = T \int d^4 x \sqrt{ G^{ind} }
\lbl{braneact}
\end{equation}
where $T$ is a dimensionfull parameter, the brane tension. 
We compute the  determinant of the metric,
\begin{equation}
G^{ind} = \det[ G^{ind}_{\mu\nu}] \ = \ e^{8\alpha \phi} \cdot \det [g_{\mu\nu}] 
\cdot \left( 1 + g_{55} e^{ 2(\beta - \alpha) \phi}
(A_\mu + \partial_\mu h) g^{\mu\nu} (A_\nu + \partial_\nu h) \right)
\lbl{5dmdet}
\end{equation}
In the limit of small 
brane fluctuations $\partial_\mu h$ we then get
from (\ref{braneact}),  (\ref{5dmdet}) the following (low energy) brane action, 
\begin{equation}
S_{brane} \ = \ T \int d^4x \sqrt{g } \, e^{4\alpha \phi} \left(
1 + \frac{1}{2} e^{2(\beta - \alpha)\phi} g_{55} (A_\mu + \partial_\mu h) g^{\mu\nu}
(A_\nu + \partial_\nu h) + ... \right)
\lbl{expact}
\end{equation}
Here the combination
\begin{equation}
\mathcal J_\mu = A_\mu + \partial_\mu h 
\lbl{Jh}
\end{equation}
is manifestly invariant under the reparametrizations (\ref{repa}).
When we implement in  (\ref{braneact}), (\ref{expact}) the changes of variables that took us from 
(\ref{HKK1}) to (\ref{HKK2}) we get the combined action
\[
S_{} + S_{brane} =  \int \! d^4x \sqrt{g } \left[ \frac{1}{16 \pi G} \left\{
R - \nabla_\mu \chi \nabla^\mu\chi \right\} \right.
\]
\begin{equation}
+ \frac{1}{4} \frac{g_{55}}{16 \pi G} e^{ \sqrt{6} \chi} \mathcal F_{\mu\nu}\mathcal F^{\mu\nu} + \frac{1}{2} T g_{55} e^{-\sqrt{\frac{2}{3}}\chi}
\mathcal J_\mu \mathcal J^\mu  \left. + \, T e^{-\sqrt{\frac{8}{3}} \chi} - \frac{2\Lambda}{16\pi G}  \,  e^{- \sqrt{\frac{2}{3}} \chi} \right]
\lbl{Stot}
\end{equation}
Here
\begin{equation}
\mathcal F_{\mu\nu} = \partial_\mu \mathcal J_\nu - \partial_\nu \mathcal J_\mu -  [\partial_\mu , \partial_\nu ] h
\label{htheta}
\end{equation}
and its last term vanishes whenever the brane fluctuations $h(x)$ are twice continuously differentiable.
It is also notable that all $\alpha$ and $\beta$ dependence has again disappeared.

Let us consider  the dilaton potential term in (\ref{Stot}),
\[
V(\chi) = T e^{-\sqrt{\frac{8}{3}} \chi} - \frac{ 2\Lambda}{16\pi G}\,  e^{- \sqrt{\frac{2}{3}} \chi} 
\]
We observe that the presence of the brane has introduced an additional term that allows us to stabilize the dilaton, 
we now have a nontrivial local minimum at
 \[
 \chi_{min} = - \sqrt{\frac{3}{2}} \ln \left[ \frac{  \Lambda}{16\pi GT}\right]
 \]
 Consequently if we redefine 
 \begin{equation}
 \mathcal J_\mu \ \to \ e \mathcal J_\mu
 \label{e}
 \end{equation}
 where
 \[
 e = \frac{1}{16\pi G} \sqrt{ \frac{ \Lambda^{3}}{g_{55} T^3} }
 \]
 and define 
 \[
 m^2 =  \frac{g_{55} \Lambda }{16\pi G}
 \]
 and redefine
 \[
 \frac{\Lambda^2}{16\pi GT}\, \to \, \Lambda
\]
then at the local minimum of the dilaton potential
the kinetic term for $\mathcal J_\mu$ acquires its correct canonical normalization and the action (\ref{Stot}) becomes
\begin{equation}
S(\chi_{min}) = \int \! d^4x \sqrt{g } \left[ \frac{1}{16\pi G} (R - 2 \Lambda)
+ \frac{1}{4}\mathcal F_{\mu\nu}\mathcal F^{\mu\nu} + \frac{m^2}{2} \mathcal J_\mu \mathcal J^\mu \right]
\lbl{finahm}
\end{equation}
Notice in particular  that we can simultaneously
have a large "Planck's mass", a small "cosmological constant", and a gauge vector mass 
that is independent of the other two, thus  at this level we avoid delicate fine tuning problems.
There is also no primordial Higgs field even though the vector boson has acquired a mass, and even
though the mass appears from dilaton interaction term
when the dilaton field acquires its classical ground state value the mechanism is different
from the Higgs mechanism. In particular the local ground state of the dilaton potential is not degenerate.

We now show that in a locally conformally flat space-time with metric 
 \begin{equation}
 g_{\mu\nu} = \frac{\rho^2}{\kappa^2}\delta_{\mu\nu}
 \label{grho}
 \end{equation}
 the action (\ref{finahm}) {\it exactly} coincides with that of the  Abelian Higgs Model in $\mathbb R^4$, in its 
 spontaneously broken phase. We have here chosen $\rho$ to have the dimension of mass, and we have  introduced an {\it 
a priori} arbitrary mass parameter $\kappa$ to ensure that the components of the metric tensor $g_{\mu\nu}$ are 
dimensionless.   

We start from  the  3+1 dimensional Abelian Higgs multiplet, a
complex scalar field $\varphi$ and 
a $U(1)$ gauge field  $A_\mu$.  We introduce a change of variables 
to another set of six independent fields $J_\mu$, $\rho$ and $\theta$ 
\begin{equation}
\begin{array}{lcl}
\varphi \  & \leftrightarrow & \ \rho \cdot e^ {i \theta } \\
A_\mu \ & \leftrightarrow & J_\mu = \frac{i}{2 e
\rho^2}\left[ \varphi^* ( \partial_\mu - i e A_\mu ) \varphi - c.c. \right]
\end{array}
\lbl{asu}
\end{equation}
This change of variables 
is invertible whenever $\rho \not= 0$, the
Jacobian is $\rho$.  When we introduce a $U(1)$ gauge transformation that acts on $\varphi$ and $A_\mu$
in the usual way, the
fields $J_\mu$ and $\rho$ are $U(1)$ gauge invariant; the vector $J_\mu$ is known as the {\it supercurrent } in applications
to superconductivity. In terms of these variables
the familiar $U(1)$ gauge invariant classical action of the 
Abelian Higgs model
\begin{equation}
{\mathcal S} = \int \! d^4x \, \left \{ \frac{1}{4} F_{\mu\nu} + |(\partial_\mu - i e A_\mu)
\varphi|^2 + \lambda |\phi|^4  \right \}
\lbl{H1}
\end{equation}
becomes
\begin{equation}
{\mathcal S} = \int \! d^4x \, \left \{  \frac{1}{4} \left(
J_{\mu\nu} - \sigma_{\mu\nu} 
\right)^2 
\!\!
+ (\partial_\mu \rho)^2 +  e^2 \rho^2 J_\mu^2
+ \lambda \rho^4 \right \}
\lbl{H2}
\end{equation}
where
\[
J_{\mu\nu} = \partial_\mu J_\nu - \partial_\nu J_\mu
\]
and the distribution
\be
\sigma_{\mu\nu} = 
\frac{1}{e}\, [\partial_\mu , \partial_\nu ] \theta
\lbl{ds1}
\ee
is the string tensor that describes vorticity, in line with the second term in (\ref{htheta}).  Its support 
in $\mathbb R^4$ coincides with the 
world-sheets of vortex cores. Except for $\theta$ in
(\ref{ds1}) there are no gauge dependent variables present in the action (\ref{H2}). Furthermore, if  gauge transformations
entail only at least twice continuosly differentiable functions, (\ref{ds1}) is gauge invariant.
Thus in the absence of (singular) vortex configurations we have the remarkable result
that the $U(1)$ gauge dependence of the Abelian Higgs Model can be entirely removed by a mere change of variables. 
For this, there is {\it no} need to introduce any  fixing of gauge nor any kind of symmetry 
breaking mechanism by the Higgs field \cite{shapo}-\cite{lude}.  We shall see that this persists in the case of the 
Weinberg-Salam Model.

As in   \cite{fad} we identify the variable $\rho$ in (\ref{H2}) with the conformal scale of a metric tensor like (\ref{grho}).
With this metric tensor we can then write  \cite{fad} the classical action (\ref{H2}) in  the following manifestly generally 
covariant form,
\begin{equation}
S  \ = \ \int d^4x \sqrt{g } \left\{ \frac{1}{16\pi G} (R + 2 \Lambda)
+ \frac{1}{4} g^{\mu\nu} g^{\lambda \eta} J_{\mu\lambda} J_{\nu\eta} + e^2 \kappa^2
g^{\mu\nu}J_\mu J_\nu \right\} 
\lbl{ahg}
\end{equation}
We have here introduced  
\begin{equation}
G = \frac{3}{8\pi \kappa^2}
\label{parG}
\end{equation}
and 
\begin{equation}
\Lambda = 3 \kappa^2 \lambda
\label{parL}
\end{equation}
and for simplicity we include the 
string tensor (\ref{ds1}) in the definition of $J_{\mu\nu}$.  With the present identifications (\ref{finahm}) and (\ref{ahg})
clearly coincide, as we asserted.
Furthermore, remarkably the vector field in the Abelian Higgs model has acquired a 
mass $\sqrt{2}e\kappa$ even though no explicit symmetry breaking and in particular
no Higgs effect has taken place. But the theory now
resides in an emergent space-time that is {\it different} from the $\mathbb R^4$ where the 
original Lagrangian (\ref{H1}) endures. This emergent
space-time dissolves away when $\rho$ vanishes. In particular,
on the world-sheet of an Abrikosov type vortex 
where (\ref{ds1}) is nontrivial we must have $\rho=0$, otherwise the energy diverges.
The metric (\ref{grho}) then vanishes and the curvature scalar 
\[
R =  - 6 \left( \frac{\kappa^2}{\rho}\right)^2 \cdot \frac{\Box \rho}{\rho}
\]
has an 
integrable singularity. As a consequence the Abrikosov vortices of the Abelian Higgs model
can be viewed  as space-time singularities in the emergent, locally conformally flat space-time.

\vskip 0.3cm
We emphasize  that we have not included any bare Higgs mass  term in (\ref{H1}).  If we include
a bare Higgs mass in (\ref{H1}) it spoils the manifestly covariant form of (\ref{ahg})
and we loose the relation between the generally covariant (\ref{finahm}) and (\ref{H1}).
While at the level of (\ref{H1}), (\ref{ahg}) the general covariance can be viewed as a pure coincidence,
at the level of (\ref{finahm}) general covariance is a symmetry of the theory and adding a term
corresponding to a bare Higgs mass in(\ref{H1}) explicitely break general covariance. If we accept a point of view 
that there is something deeper between (\ref{finahm}) and  (\ref{H1})  than pure coincidence, there
can not be any bare Higgs mass in (\ref{H1}) either. Consequently, if the theory (\ref{H1}) 
is regularized and quantized in a manner that continues to relate it to (\ref{finahm}) and 
thus respects the generally
covariant interpretation (\ref{ahg})  {\it no} bare Higgs mass term can emerge as a quantum correction, Higgs mass
is forbidden by an appropriate Ward-Takahashi identity
that reflects  the underlying generally covariant interpretation, and the relation between (\ref{H1}) and (\ref{finahm}).
We may say that the Higgs mass is subject to  {\it dimensional censorship. } 

 In the sequel we shall show that the present construction persists in the non-Abelian context
of Weinberg-Salam model, thus such a dimensional censorship might be a natural resolution to the electroweak hierarchy problem: 
A mass term for the Higgs field is not consistent with the symmetries of the theory as these symmetries have their origin in higher 
dimensional general covariance and a bare Higgs mass  breaks this general covariance.

However, this does not prevent $\rho$ from
having a non-trivial ground state value: Even when the perturbative contributions  that are not consistent with the interpretation
in terms of general covariance
are removed,  it has been argued 
in \cite{guba1}-\cite{fad2} both on general grounds and using numerical lattice
simulations that in a quantum gauge theory the condensate
\begin{equation}
<J^2_\mu> \ = \ \pm \Delta^2
\label{A2}
\end{equation}
is non-vanishing. Here we have added the sign to reflect  the fact that in {\it Minkowski} space  the condensate can be either  time-like or space-like:
Following  \cite{max} we expect that there is a phase transition with order parameter
\[
< J_4^2> - < J_i^2> \ = \  \pm \Delta^2 
\]
and according to (\ref{H2}) the sign corresponds to positive {\it resp.} negative Higgs mass  
\[ 
e^2 J_\mu^2 \rho^2 \  \to \ e^2 <J_\mu^2> \rho^2 \ \sim  \  \pm e^2 \Delta^2 \rho^2 \  =  \ \pm e^2 \Delta^2 |\phi |^2
\]
that is  
the phase transition is between the symmetric and broken Higgs phases in the conventional parlance. 
Indeed, it has been proposed that the quantity (\ref{A2}) determines a natural and {\it gauge invariant} \cite{guba1}-\cite{fad2} 
dimension-two condensate in a gauge theory.  From (\ref{H2}) we then estimate that  
in the London limit where $\rho = \rho_0$ is a constant corresponding  to the conventional 
situation where the Higgs field is in a translationally invariant ground state,  we have the non-vanishing condensate value
\begin{equation}
\rho_0^2 = \frac{e^2 \Delta^2}{\lambda}
\label{A2b}
\end{equation}

Finally we comment on the following: As such it should not come as a surprise that a Poincare invariant  field theory can be written in 
a generally covariant form. For this all one needs is to implement a transformation from the Cartesian 
to a generic coordinate system, the result always has a generally covariant form. However, the {\it peculiarity} in the present case is 
that now the metric  tensor is  constructed from one of the field variables, and  that the Hilbert-Einstein action makes an appearance.  
In fact, it has been proposed that any unitary four dimensional field theory that possesses both  
Poincare and rigid scale symmetry is  invariant under the entire conformal group $SO(5,1)$ \cite{polch}. 
The results of \cite{poly}, \cite{sund} in the case of  (special) conformally invariant $\lambda \phi^4$   
and of \cite{fad2}, \cite{fad},  \cite{lude} in the case of  (special) conformally invariant
Yang-Mills-Higgs theories then suggest that the $SO(5,1)$  special conformal symmetry 
in $\mathbb R^4$ can be extended to include invariance under local conformal 
transformations, and the ensuing theory can be cast in  a manifestly generally covariant form
with the conformal scale constructed from the field variables \cite{foot1}.

\section{Supercurrents And The Weinberg-Salam Model}
\vskip 0.5cm

We now proceed to   the non-Abelian $SU_L(2) \times U_Y(1)$ invariant 
Weinberg-Salam model, where our goal is to relate its symmetry structure with that of a higher dimensional gravity
theory to address the electroweak hierarchy problem. The present Section describes how the pertinent Lagrangian
(\ref{ahg}) is derived, and the remaining Sections are devoted to relate the ensuing (accidental?) general covariance
to a higher dimensional local symmetry.
 
The bosonic part of the Weinberg-Salam  Lagrangian is
\begin{equation}
\cL_\WS
= \frac{1}{4} \vec G_{\mu\nu}^2(A) + \frac{1}{4} F^2_{\mu\nu}(Y)
+ | D_\mu { \Phi}|^2 + \lambda |\Phi|^4 
\lbl{su2L}
\end{equation}
We work in a flat spacetime with Euclidean signature, and follow  the notation of \cite{abers}:
The
matrix-valued $SU_L(2)$ isospin gauge field is
\[
\widehat A_\mu \equiv  A^a_\mu \tau^a = {\vec {A}}_\mu \cdot \vec \tau 
\]
with $\tau^a$ the isospin Pauli matrices, and $Y_\mu$ is the Abelian $U_Y(1)$ hypergauge field. 
The field strengths are
\be
\vecG_{\mu\nu}(A) & = & \partial_\mu \vec A_\nu -
\partial_\nu \vec A_\mu - q \, \vec A_\mu \times \vec A_\nu\,,
\lbl{eq:G}\\
F_{\mu\nu}(Y) & = & \partial_\mu Y_\nu - \partial_\nu Y_\mu\, .
\lbl{eq:F}
\ee
and the $SU_L(2) \times U_Y(1)$ covariant derivative is
\begin{equation}
D_\mu = \bbbone \, \partial_\mu - i \frac{q}{2}
\widehat A_\mu  - i \frac{q'}{2} Y_\mu\,\bbbone\,,
\lbl{covdev}
\end{equation}
where $\bbbone$ is the $2 \times 2$ unit matrix in the isospin space.

Notice  that  as in the case of Abelian Higgs model, we do {\it not} add any (bare) mass term to the  complex isospinor Higgs boson
$\Phi$, the Higgs potential is conformally invariant, of the Coleman-Weinberg form. It turns
out that as in the Abelian case there is no need for a conventional kind of a Higgs effect. Instead, the gauge boson 
masses will emerge at the classical level simply from a change of variables in a combination with a geometric interpretation, 
while the modulus of the Higgs field acquires a ground state expectation value from a non-Abelian generalization  of (\ref{A2}).

We start by generalizing the construction of the gauge invariant supercurrent (\ref{asu}) 
to the case of the Weinberg-Salam model. We  follow largely the approach  in \cite{fad}, 
with some minor changes that are convenient when we proceed to generalize the results of Section 2. 

We start by  decomposing the Higgs field $\Phi$ as follows,
\begin{equation}
\Phi \ = \phi \, \mathcal X \ \ \ \ {\rm with}
\quad \ \ \
\phi = \rho \, e^{i\theta} \
\ \ \ \& \ \ \ \mathcal X = \
\mathcal U
\left( \begin{matrix} 0 \\ 1 \end{matrix} \right)
\lbl{phi}
\end{equation}
Here $\phi$ is a complex field,  $\mathcal X$
a two-component complex isospinor with $|\mathcal X | = 1$, and we take $\mathcal U $ to 
be a $2\times 2$ $SU(2)$ matrix.  The $SU_L(2) \times U_Y(1)$
gauge transformation
acts on $\Phi$ as follows,
\begin{equation}
\Phi \ \to \ e^{i \omega_Y} \Omega \, \Phi
\ \  \Rightarrow \ \ \left\{ \begin{matrix}
\phi \hskip 2mm \ \longrightarrow e^{i \omega_Y} \phi
\\
\mathcal X \longrightarrow \ \Omega \mathcal X \end{matrix} \right.
\lbl{eq:splitting}
\end{equation}
where $\Omega \in SU_L(2)$ and $e^{i\omega_Y} \in U_Y(1)$.
The decomposition (\ref{phi}) also introduces a new (internal) {\it compact} gauge group
\begin{equation}
U_\inter(1):\quad \begin{matrix}
\phi \to e^{- i \omega_c} \phi \\
\mathcal X \to e^{i \omega_c} \mathcal X \end{matrix}
\lbl{eq:internal}
\end{equation}
which leaves the field $\Phi$ intact.
Note that the spinor $\mathcal X \equiv \mathcal X_1$ and its
isospin conjugate 
\[
\mathcal X_2 = i \tau_2 { {\mathcal X}}_{}^*
\] 
form an orthonormal basis
($i,j=1,2$ and $a,b = \uparrow, \downarrow$),
\[
\mathcal X_i^\dagger \cdot \mathcal X_j \
\equiv \sum_{a = \uparrow, \downarrow} \mathcal X_{ia}^* \mathcal X_{aj}
\ = \ \delta_{ij} 
\]
\[
\sum_{i=1,2}
\mathcal X_{i a}^{} \mathcal X_{ib}^\dagger \ = \ \delta_{ab}
\]

When we introduce the conjugate Higgs field
\[
\Phi_c = \phi \mathcal X_2
\] 
we find the $SU_L(2)\times U_Y(1)$ supercurrents $(J^\pm_\mu, J^3_\mu)$ 
and $\mathcal Y_\mu$  (with $J_\mu^\pm 
= J^1_\mu \pm i J^2_\mu$)  by  expanding the covariant derivative of the Higgs field
in the spinor basis ($\mathcal X_1,\mathcal X_2$) \cite{fad}
\begin{equation}
D_\mu \Phi \ = \ \Bigl[\frac{1}{\rho}\partial_\mu \rho - \frac{i}{2}
\Bigl( q J_{\mu}^3 - q'  \mathcal Y_\mu  \Bigr) \Bigr] \, \Phi
+ i \frac{q}{2} J^+_\mu \cdot \Phi_c
\lbl{DPhi}
\end{equation}
Explicitely, 
\begin{eqnarray}
J_\mu^+ \! & = & \! - \frac{2i}{q}\mathcal X_{2}^\dagger \!
\left( \partial_\mu + \frac{iq}{2} \vec{A}_\mu \cdot \vec{\tau} \right) \! \mathcal X_1^{}
\equiv
\vec A_\mu \cdot \vec e_+ + \frac{i}{q} \vec e_3 \cdot \partial_\mu \vec e_+\,,
\lbl{J+}\qquad \\
J^3_\mu \! & = & \! - \frac{2i}{ q}
\mathcal X_1^\dagger
\! \left( \partial_\mu + \frac{iq}{2} \vec A_\mu \cdot \vec{\tau} \right) \!
\mathcal X_1^{}
\equiv
\vec A_\mu \cdot \vec e_3 - \frac{i}{2q} \vec e_- \cdot
\partial_\mu \vec e_+ \, ,
\lbl{J3}
\end{eqnarray}
and
\begin{equation}
\mathcal Y_\mu \  =  \  \frac{i}{q' |\phi|^2}
\Bigl[ \phi^\star \Bigl(\partial_\mu - i
\frac{q'}{2} Y_\mu \Bigr)\phi - c.c. \Bigr] 
\lbl{j}
\end{equation}
and $\vec e_i  \ (i=1,2,3)$ are three mutually
orthogonal unit vectors defined by
\begin{eqnarray}
\vec e_3 & = & - \frac{\Phi^\dagger {{\vec\tau}} \Phi}{\Phi^\dagger \Phi}
         \equiv - \mathcal X_1^\dagger {{\vec\tau}} \mathcal X_1 \\
\vec e_+ & = &  \vec e_1 + i \vec e_2 = \mathcal X_2^\dagger {{\vec\tau}} \mathcal
X_1
\lbl{eX}
\end{eqnarray}
The $SU(2)$ matrix $\mathcal U$ in (\ref{phi}) combines these into
\begin{equation}
\vec e_i \, \tau^i = \mathcal U^{-1} \vec \tau \, \mathcal U
\lbl{eU}
\end{equation}  
and the $3\times 3$ matrix ${e_i}^a$ is an element of $SO(3)$ since
\[
{e_i}^a {e_j}^a = \delta_{ij} \ \ \ \ \ \& \ \ \ \ \ {e_i}^a {e_i}^b = \delta^{ab}
\]

We view (\ref{J+})-(\ref{eX}) as the following change of variables,
\begin{equation}
(\vec A_\mu, Y_\mu, \Phi) \to (  J^3_\mu , J^\pm_\mu , \mathcal Y_\mu, \vec e_i , \rho )
\lbl{chav}
\end{equation}
On both sides of (\ref{chav}) there are sixteen real fields, and (\ref{chav}) 
is an invertible change of variables whenever $\rho \not= 0$;  the
Jacobian is $\rho^3$. When we substitute (\ref{chav}) in (\ref{su2L}) we get \cite{fad}
\begin{eqnarray}
& &
\hskip -4mm
\cL_\WS =  (\partial_\mu \rho)^2 \!  + \lambda \hskip 0.6mm \rho^4 \ + \ 
\frac{1}{4} \left( \! \vec G_{\mu\nu}(\vec J)
+ \frac{4 \pi}{q} {\vec{\widetilde \Sigma}}_{\mu\nu} \!
\right)^2 \! \! \! + \frac{1}{4} \left( \!
F_{\mu\nu}(\mathcal Y)
+ \frac{4\pi}{q'} \widetilde \sigma_{\mu\nu}^\phi \!
\right)^2
\nonumber\\
& & \hskip 3.0cm
+ \frac{\rho^2}{4}\left( q J^3_\mu - q' \mathcal Y_\mu\right)^2
\! + \frac{\rho^2 q^2}{4} J_\mu^+ J_\mu^- 
\lbl{wsgi}
\end{eqnarray}
Here $\vec G_{\mu\nu}$ and $F_{\mu\nu}$ are
the field strength tensors  of $\vec J_\mu$ {\it resp.} $\mathcal Y_\mu$,
\be
\vecG_{\mu\nu}(\vec J) & = & \partial_\mu \vec J_\nu -
\partial_\nu \vec J_\mu - q \, \vec J_\mu \times \vec J_\nu\,,
\lbl{eq:GJ}\\
F_{\mu\nu}(\mathcal Y) & = & \partial_\mu \mathcal 
Y_\nu - \partial_\nu \mathcal Y_\mu\, .
\lbl{eq:FY}
\ee
The  $\tilde\sigma^\phi_{\mu\nu}$ is the dual of the string tensor (\ref{ds1}) in the present case and
 the $\vec e_i $ appear only through the singular quantity
\begin{equation}
\Sigma_{\mu\nu}^i = \frac{1}{8\pi} \epsilon^{ijk} ( {\vec e}^{\hskip 0.5mm j} 
\cdot [\partial_\mu , \partial_\nu] \vec e^{\hskip 0.5mm k})
\lbl{schoen}
\end{equation}
which is a non-Abelian generalization of  (\ref{ds1}). 
\vskip 0.3cm
\noindent
We make the following two remarks:

\vskip 0.4cm

1) If we
resolve the relations (\ref{J+}), (\ref{J3}) for $A_\mu^i$ we 
can combine (\ref{J+}), (\ref{J3}) into
\[
J^a_\mu = A_\mu^i {e_i}^a + \frac{1}{2q}\epsilon^{abc} {e^j}_{b} \partial_\mu {e_{jc}}
\]
and when we invert this by using the fact that ${e^i}_a \in SO(3)$ we get
\begin{equation}
A_\mu^i = {e^i}_a J_\mu^a + {e^i}_{a}  
\frac{1}{2q} \epsilon^{abc}  {e^j}_b \partial_\mu {e_{jc}}
= {e^i}_a \{ J_\mu^a + \frac{1}{2q} \epsilon^{abc} {e^j}_b \partial_\mu
{e_{jc}}\}
\lbl{AinJ}
\end{equation} 
Here the second term is a pure gauge {\it i.e.} left-invariant Maurer-Cartan form,
\begin{equation}
(\epsilon^{abc} {e^j}_{b}\partial_\mu {e_{jc}} )\cdot {e^i}_{a}\,  \frac{\tau^i}{2i} = 
\mathcal U^{-1} \partial_\mu 
\mathcal U
\lbl{pur}
\end{equation}
where $\mathcal U \in SU(2)$ is defined in (\ref{phi}), (\ref{eU}).

\vskip 0.4cm

2) Following (\ref{A2}) and \cite{guba1}-\cite{fad2} and \cite{max}  we propose that in the
quantum theory the expectation values
\begin{equation}
< (q J^3_\mu - q' \mathcal Y_\mu)^2 \! > \ = \ \pm \Delta_3^2
\label{Zakh1}
\end{equation}
\begin{equation}
< q^2 J^+_\mu J^-_\mu \! > \ = \ \pm \Delta^2_\pm
\label{Zakh2}
\end{equation}
are non-vanishing, with the sign (in Minkowski space) depending on whether the condensate is space-like or time-like.
From (\ref{wsgi}) we then estimate for the ground state value $\rho_0$ of the Higgs modulus
\[
\rho_0^2 \ = \  \frac{1}{4\lambda} ( \pm \Delta_3^2 \pm \Delta_\pm^2 )
\]
As in the Abelian case we again conclude that even though there is no bare Higgs mass, a non-vanishing Higgs condensate can
be generated by the condensation of the intermediate vector bosons. Furthermore, the sign of the condensate {\it i.e.} whether we
are in the broken or symmetric Higgs phase depends on the signs of the condensates (\ref{Zakh1}), (\ref{Zakh2}) that is whether
we have a time-like or space-like condensate in the Minkowski space \cite{max}.

\vskip 0.4cm

In  line with (\ref{ahg})  we can interpret the Lagrangian (\ref{su2L}) in terms of
local conformal geometry. As in  (\ref{grho}) we identify $\rho$ with the
conformal scale of a metric tensor, and repeating the steps that
led to (\ref{ahg}) we get  \cite{fad}
\begin{equation}
\cL_\WS \ = \ \sqrt{ G} \left\{ \frac{1}{16\pi G}(
R + 2 \Lambda) + \cL_{M} \right\} 
\lbl{egw1}
\end{equation}
where  the matter Lagrangian $\cL_M$ is
\begin{equation}
\cL_{M}  =   \frac{1}{4} 
{\vec G}_{\mu\nu} \cdot {\vec G}^{\mu\nu} + \frac{1}{4}
{F}_{\mu\nu} {F}^{\mu\nu}
+ \kappa^2 (q^2 + {q'}^2)
Z_\mu Z^\mu +
\kappa^2 q^2
W_\mu^+ W^{\mu -} 
\lbl{ewg2}
\end{equation}
and the indices are raised and lowered using the metric tensor (\ref{grho}).
As in (\ref{parG}), (\ref{parL}) we have here introduced  
\[
G = \frac{3}{8 \pi \kappa^2}
\]
and 
\[
\Lambda = 3 \kappa^2 \lambda 
\]
and 
the $SU_L(2)\times U_Y(1)$ invariant $W$--bosons are
$W^\pm_\mu = J^\pm_\mu$, while the $Z$--bo\-son and photon $A_\mu$ are
\begin{eqnarray}
Z_\mu & = & \cos\theta_W\, J^3_\mu - \sin\theta_W\, \mathcal Y_\mu \,, \\
A_\mu & = & \sin\theta_W\, J^3_\mu + \cos\theta_W\, \mathcal Y_\mu\,,
\lbl{eq:ZA}
\end{eqnarray}
where $\theta_W$ is the Weinberg angle, it has the experimental  the low momentum transfer value \cite{E158}
\begin{equation}
\sin^2\theta_W = \frac{{q'}^2}{ q^2 + {q'}^2 } \ = \ 1 - \frac{ M^2_W }{M^2_Z}  \approx \ 0.2397 \pm 0.0014
\label{lowm}
\end{equation}
By recalling the  (low energy) Thomson limit value 
\[
\alpha = \frac{e^2}{4\pi} \ \approx \ \frac{1}{137}
\]
for the electric charge  this gives for the $SU_L(2)\times U_Y(1)$ couplings the following numerical values
\begin{equation}
e \ =  \ q \sin \theta_W \ = \ q' \cos \theta_W \  \   \  \Rightarrow \  \ \ q \approx \ 0.619 \ \ \& \ \ q' \approx 0.312
\label{expg}
\end{equation}

The Lagrangian (\ref{egw1}), (\ref{ewg2})  has the familiar form of the spontaneously broken
Einstein-Weinberg-Salam Lagrangian. It describes  the conventional electroweak interactions of the massive $W$ and $Z$ bosons
in a conformally flat space-time, that becomes a flat $\mathbb R^4$ in the London limit where $\rho$ is a constant.
As in the Abelian case,  we find it notable that now it is the dimensionfull parameter  $\kappa$ that gives rise to the vector 
masses, {\it not} the Higgs ground 
state expectation value as in  conventional approach. In fact, in line with the Abelian case we immediately observe that the presence of a bare
Higgs mass term would not allow us to write the Lagrangian  in the generally covariant form.  While at the present level of argumentation
this generally covariant form could be viewed as accidental, in the rest of this paper  we argue that it may also reflect dimensional
censorship imposed by an underlying higher dimensional
structure.   If so,  the addition of a bare Higgs mass would lead to an explicit breaking of
general covariance. But if the quantization is performed in a manner that respects  general covariance 
 as a Ward-Takahashi like identity, a Higgs mass term can not appear and the electroweak hierarchy 
problem may have a simple resolution. Furthermore,  the conventional Higgs field becomes  metamorphosed  into 
the local conformal scale,    and as in the Abelian case in the quantum theory its modulus may acquire a non-vanishing expectation value from the 
condensation of the intermediate vector bosons, without violating the underlying general covariance.

\section{Squashed three-sphere}
\label{s7s}

\vskip 0.5cm

We  now proceed to disclose how  the symmetry structure of the Weinberg-Salam Lagrangian in its representation (\ref{egw1}), (\ref{ewg2}) 
becomes embedded in  the brane world, to reflect the potential presence of higher dimensions. Following Section \ref{vmfb}  we start with 
a Kaluza-Klein setup  which we build on  $\mathbb M^4 
\times \mathbb S^3$. Here $\mathbb M^4$ is the space-time four-manifold with metric components
$g_{\mu\nu}$ ($\mu,\nu = 0,1,2,3$). Eventually we shall specify to a locally conformally flat
space-time, to reproduce the result (\ref{egw1}), (\ref{ewg2}). 
The internal  $\mathbb S^3 \sim SU(2)$ is the gauge group manifold that we eventually
squash.  It turns out that the squashing parameter will allow us to 
relate  the gauge couplings $q$ and $q'$  in the Weinberg-Salam model.

In this Section we present some useful
relations for   $SU(2)\simeq  \mathbb S^3$, both with the standard metric  and its squashed generalization. The results are largely
familiar  \cite{pope} but there are  some new details.
We describe the manifold $SU(2) \simeq \mathbb S^3$ in terms of  the $2\times 2$ matrix $\mathcal U  $  that we introduced 
in (\ref{eU}). For concreteness we use the following explicit Euler angle parametrization
\begin{equation}
\mathcal U = -i \left( \begin{matrix} \sin \frac{\theta}{2} e^{\frac{i}{2}\phi_+} &
- \cos \frac{\theta}{2} e^{ \frac{i}{2} \phi_-} \\
- \cos \frac{\theta}{2} e^{- \frac{i}{2} \phi_-} & - \sin \frac{\theta}{2} e^{-\frac{i}{2}
\phi_+} \end{matrix} \right)
\lbl{UEu}
\end{equation}
where  $0 \leq \theta \leq \pi$ and
$0 \leq \phi_{\pm} \leq 2\pi$ are local coordinates on $\mathbb S^3$.
The natural metric $g_{mn}$ ($m,n = 1,2,3$)  on $\mathbb S^3$ is
the bi-invariant Killing two-form,
\begin{equation}
ds^2 = 2 \, Tr( d\mathcal U d\mathcal U^{-1} ) = g_{mn} d\vartheta^m d\vartheta^n 
=  (d \theta)^2 + \sin^2\!  \frac{\theta}{2} \,  (d\phi_+)^2 + \cos^2 \! \frac{\theta}{2}  \, (d\phi_-)^2
\label{bim}
\end{equation}
We write the left-invariant Maurer-Cartan one-form (\ref{pur}) as follows,
\begin{equation}
\mathcal U^{-1} d \mathcal U  = {L^a_{m}} d\vartheta^m \frac{1}{2i} \tau^a
\lbl{mc2}
\end{equation} 
where $\tau^a$ are the Pauli matrices.  The right-invariant Maurer-Cartan is
\begin{equation}
\mathcal U d \mathcal U^{-1} = R^a_m d\vartheta^m \frac{1}{2i} \tau^a
\label{mc2r}
\end{equation}
The components ${L^a_{m}}$ and ${R^a_{m}}$  can both be identified as the dreibeins for the metric (\ref{bim}),
\begin{equation}
g_{mn} = \delta_{ab} {L^a_{m}}  {L^b_{n}}  \ = \ \delta_{ab}{R^a_{m}} {R^b_{n}}
\lbl{gGG}
\end{equation}
The one-forms ${L^a}  = {L^a_m} d\vartheta^m$ and $R^a = R^a_m d\vartheta^m $ are also 
subject to the $SU_L(2)$ Maurer-Cartan equation, {\it e.g.}
\begin{equation}
d{L^a}  = - \frac{1}{2} \epsilon^{abc} {L^b}  \wedge {L^c} 
\lbl{dg}
\end{equation}
and explicitely we have 
\begin{eqnarray}
{L^1}   & = & {e_3}^1  d \psi_+ -  {e_2}^1 \, d\theta 
\lbl{lhmc1}
\\
{L^2}  & = & {e_3}^2   d\psi_+ - {e_2}^2  \, d\theta 
\\
{L^3}  & = &  {e_3}^3 d\psi_+ - \  d\psi_- 
\lbl{lhmc3}
\end{eqnarray}
where we have defined
\begin{equation*}
\psi_{\pm} = \frac{1}{2} (\phi_+ \pm \phi_-) 
\end{equation*}
and we have introduced the right handed orthonormal triplet (\ref{eU}),
\begin{equation}
\vec e_1 \ = \ \left( \begin{matrix} \cos \psi_- \cos \theta \\ \sin \psi_- \cos\theta \\ - \sin\theta \end{matrix} \right) \ \ \ \ \ \  \& \ \ \ \ \ 
\vec e_2 \ = \ \left( \begin{matrix} - \sin \psi_-   \\ \cos \psi_-  \\ 0 \end{matrix} \right) 
\ \ \ \& \ \ \  \ \  \vec e_3 \ = \    \left( \begin{matrix} \cos\psi_- \sin \theta 
\\ \sin \psi_- \sin\theta \\  \cos\theta \end{matrix} \right) 
\lbl{e1e2n}
\end{equation}
The ensuing explicit  realizations of the right Maurer-Cartan one-forms 
are obtained simply by sending 
\begin{equation}
(\theta, \psi_+ , \psi_-)  \to - (\theta, \psi_-, \psi_+)
\label{rlm}
\end{equation}
There are  three left-invariant Killing vector fields 
\[
K_L^a = (K_L^a)^m \frac{ \partial}{\partial \vartheta^m} \ \ \ \ \ (m=1,2,3)
\]
that can be identified as the canonical duals of the one-forms $L^a$.  
With (\ref{lhmc1})-(\ref{lhmc3}) this gives us the explicit realization
\begin{equation}
K_L^1  = \left\{ \sin \psi_- \partial_\theta +  \cos\psi_- \cot \theta 
\partial_{\psi_-} \right\} \ + \ \frac{\cos\psi_-}{\sin\theta} \partial_{\psi_+} \ = \ 
l^1 + t^1
\lbl{KT1}
\end{equation}
\begin{equation}
K_L^2 
= \left\{ - \cos \psi_- \partial_\theta + \sin\psi_- \cot \theta \partial_{\psi_-} \right\}  \ 
+ \ \frac{\sin\psi_-}{\sin\theta} \partial_{\psi_+} \ = \ 
l^2 + t^2
\lbl{KT2}
\end{equation}
\begin{equation}
K_L^3   =  -  \partial_{\psi_-}  \equiv  l^3
\lbl{KT3}
\end{equation}
The commutators of the Killing vectors determine a representation
of the $SU_L(2)$ Lie algebra, 
\begin{equation}
[ K_L^a , K_L^b ] = - \epsilon^{abc} K_L^c
\lbl{KKK}
\end{equation}
Furhermore, in $l^a$ we identify the standard $SO(3)$ angular momentum operators 
with
\[
[ l^a , l^b ] = - \epsilon^{abc} l^c
\]
while the $t^a$ obey the one-cocycle condition
\begin{equation}
[l^a , t^b ] + [t^a , l^b ]  =  -\epsilon^{abc} t^c 
\lbl{LTT}
\end{equation}
\begin{equation}
[t^a, t^b]  =  0
\lbl{TT}
\end{equation}

We also note the possibility to introduce a two-cocycle into the Lie algebra (\ref{KKK}). For this we deform
the Killing vectors into
\begin{equation}
K_L^a  \to \hat K_L^a \ \equiv \ K_L^a + \alpha \cdot  T_L^a \ = \  K_L^a + \alpha \cdot e_3^a \ \partial_{\psi_+}
\label{2co}
\end{equation}
The deformed Lie algebra is
\[
[ \hat K^a_L , \hat K^b_L ] \ = \ - \epsilon^{abc} \hat K^c_L + \alpha \cdot \epsilon^{abc} T_L^c
\]
and in the equivariant subspace where
\[
T_L^a F (\theta, \psi_{-} , \psi_+) \ = 0 \ \Rightarrow \ F=F(\theta,\psi_-)
\]
these deformed Killing vectors act like the original ones.

We recall that the Killing vectors generate an isometry of the metric (\ref{bim}).  With $\mathcal L_{a}$
the Lie derivative in the direction of $K_L^a$
\begin{equation}
\mathcal L_{a} g_{mn} = 0 \ \ \ \ \ \ \, i=1,2,3
\lbl{Lg}
\end{equation}
They are also orthonormal,
\begin{equation}
g_{mn} (K_L^a)^m (K_L^b)^n = \delta^{ab}
\lbl{gKK}
\end{equation}
Again, the ensuing explicit  realization of the right Killing vectors 
is obtained from (\ref{rlm}).
In particular, from (\ref{KT3}) we get
\begin{equation}
K^3_R \ \equiv R \ = \ + \partial_{\psi_+}
\lbl{KR}
\end{equation}

We can explicitely  break the $SU_L(2) \times SU_R(2)$ isometry of $\mathbb S^3$ into $SU_L(2) \times U_R(1)$ by
squashing \cite{pope} the three sphere. For this  we modify the metric tensor (\ref{bim}) into the
following one-parameter family of metrics,
\begin{equation}
 g_{mn}   d\vartheta^m d\vartheta^n \ = \ 
(d\theta)^2 + \sin^2 \! \theta \, (d\psi_-)^2 + ( d\psi_+ -  \cos \! \theta \,  d\psi_-)^2
\label{nosqu}
\end{equation}
\begin{equation}
\to (d\theta)^2 + \sin^2 \! \theta \, (d\psi_-)^2 + \xi^2 ( d\psi_+ - \cos  \! \theta \, d\psi_-)^2 = g^\xi_{mn}  d\vartheta^m d\vartheta^n
\lbl{squ}
\end{equation}
A dreibein representation of this 
squashed metric is obtained {\it e.g.} in terms of the right Maurer-Cartan one-forms by modifying them as follows
\begin{eqnarray} 
E^1 & = &  R^1 = \  e^1_3  \ d\psi_-  -  e^1_2 d\theta
\lbl{rhmca}
\\
E^2 & = &  R^2  =  \ e^2_3 \ d\psi_- - e^2_2  d\theta
\lbl{rhmcb}
\\
E^3 & = &   \xi \cdot R^3  =  \ \xi \cdot ( e^3_3 d\psi_- - d\psi_+)  
\lbl{rhmcc}
\end{eqnarray}
where we have implemented the left-right conjugation (\ref{rlm}) in the triplet (\ref{e1e2n}).
This gives the dreibein decomposition of the squashed metric tensor (\ref{squ}),
\begin{equation}
g^\xi_{mn}  = E^i_m E^j_n \delta_{ij} = R^1_m R^1_n + R^2_m R^2_n + \xi^2 R^3_m R^3_n
\label{Rdrei}
\end{equation}
Alternatively, we can introduce the following dreibein one-forms to similarly decompose  the squashed metric,
\begin{eqnarray}
E^1  & =  & \left\{ {e_3}^1 \cos \theta d\psi_-  -  {e_2}^1 d\theta \right\} + \xi \, {e_3}^1  
(d\psi_+  -  \cos\theta d\psi_- )
\lbl{lhmca}
\\
E^2 & =  &  \left\{  {e_3}^2 \cos\theta d\psi_-  -  {e_2}^2 d\theta  \right\} 
+ \xi \, {e_3}^2 ( d\psi_+ - \cos\theta d\psi_- )
\lbl{lhmcb}
\\
E^3 & =  & \left\{ {e_3}^3 \cos\theta d\psi_- - d\psi_- \right\} + \xi \, {e_3}^3 (  d\psi_+ - \cos\theta d\psi_-)
\lbl{lhmcc}
\end{eqnarray}
These dreibeins have the advantage that  in the $\xi \to 0$  limit none of them vanishes and they go smoothly over to give the standard metric 
on the two-sphere $\mathbb S^2$ with local coordinates $(\theta, \psi_-)$.   This will become convenient in Section \ref{lasts}.

Finally, we remind that for {\it any} value of the squashing parameter $\xi $ in (\ref{squ}) 
the  {\it original} left Killing vectors (\ref{KT1})-(\ref{KT3}) in addition
of the 3$^{rd}$ component of the right Killing vector (\ref{KR}) 
remain as the  Killing vectors of the squashed sphere, independently of $\xi$. 
Together they generate the Lie algebra
$SU_L(2) \times U_R(1)$. But since  the   $\psi_- \leftrightarrow
\psi_+$ symmetry becomes broken for $\xi \not=1$, the squashed three-sphere
does not anymore admit the full  right invariant  $SU_R(2)$ isometry.

\section{Weinberg-Salam And Squashed Sphere}

\vskip 0.5cm

We now generalize the derivation  of (\ref{finahm}) to inspect how the symmetry structure of 
Weinberg-Salam Lagrangian  (\ref{egw1}), (\ref{ewg2}) becomes embedded
in  the brane world.
Our starting point is the {\it pure} seven dimensional 
Hilbert-Einstein action {\it without} a cosmological constant
on the manifold $\mathbb M^4 \times \mathbb S^3_\xi$ 
\begin{equation}
S = \frac{1}{16\pi G} \frac{1}{V_\xi} \int d^4x d^3\vartheta \, \sqrt{g_{(7)}} \,  \, R_{(7)} 
\lbl{E7}
\end{equation}
We choose  $\mathbb M^4$ to be  a generic 
four-manifold with 
%(Euclidean signature) 
metric tensor $g_{\mu\nu}$ and local coordinates $x^\mu$, 
and $\mathbb S^3_\xi$ is the squashed three-sphere now with metric 
\begin{equation}
ds^2 =  \frac{ r^2}{4} g^\xi_{mn} d\vartheta^m d\vartheta^n =
\frac{r^2}{4} \{  (d\theta)^2 + \sin^2 \! \theta \, (d\psi_-)^2 + \xi^2 ( d\psi_+ - \cos \theta \, d\psi_-)^2 \}
\lbl{squ2}
\end{equation}
We take $r$ to be a constant so that the volume  of  the squashed sphere is 
\[
V_\xi = 2\pi^2 \xi r^3
\]
We introduce the following Kaluza-Klein decomposed metric 
over $\mathbb M^4 \times \mathbb S^3_\xi$
\[
ds^2 = g_{\alpha \beta} dy^\alpha dy^\beta 
\]
\begin{equation} 
= g_{\mu\nu} dx^\mu dx^\nu
+ \frac{r^2}{4} g^\xi_{mn} \{ d\vartheta^m +  K_L^{am} A^a_\mu dx^\mu +  R^m B_\mu dx^\mu\}
\{d\vartheta^n + K_L^{bn} A^b_\nu dx^\nu + R^n B_\nu dx^\nu\}
\lbl{g7}
\end{equation}
Here  $K_L^{am}$ are the components
of the left Killing  vectors (\ref{KT1})-(\ref{KT3}) and $R^m \equiv K_R^{3m}$ are the components of  (\ref{KR}).

At this point we  note the following:
The decomposition (\ref{g7}) is {\it not} the most general one of the metric tensor, in particular it
does not include the higher dimensional dilaton fields \cite{cho}, \cite{sergey}. However, 
here the goal is {\it not} to deduce the Weinberg-Salam model from a higher dimensional gravity
theory using the Kaluza-Klein approach, this remains a problem that still waits for an elegant solution.
Instead, as explained in the introduction we search for a resolution to the Higgs mass hierarchy problem
from symmetries of the Weinberg-Salam Lagrangian that have a higher dimensional interpretation.
We only inquire what  is the minimal amount of higher dimensional structure that we need to look into, 
in order to address and  hopefully resolve the electroweak hierarchy problem in the limited context of 
the bosonic Weinberg-Salam Lagrangian.   In particular, how can we argue that the generally covariant
form (\ref{egw1}), (\ref{ewg2}) is not just an accidental coincidence that does not need to survive quantization, 
but a reflection of dimensional censorship imposed by an 
inherent higher dimensional  symmetry structure that may be at the root of solving the 
electroweak hierarchy problem.

It is  natural to assume that the higher dimensional manifold has the (local) product structure of  a 
four-dimen\-sional space-time  manifold $\mathbb M_4$ with some  internal manifold.  Furthermore, whatever the
structure of the internal manifold  it should somehow relate to the  squashed three-sphere,  
since the commutators of its  Killing vectors coincide  with the Lie algebra of $SU(2) \times U(1)$  \cite{pope}. 
Consequently  we do not think it is unreasonable to assume that {\it any} approach to electroweak interactions that 
involves a higher dimensional construct,  somehow  relates to the structure of  a squashed three sphere as an internal
manifold, and for this
reason we here select it as our internal manifold.

Note  that as in the Abelian case we discussed in Section II, in a complete and fully consistent  Kaluza-Klein approach
where the dilaton fields are included we would arrive at an extension of the Weinberg-Salam model with 
additional scalar fields that are due to the dilatons \cite{cho}, \cite{sergey}.  This can  be of importance  {\it if} LHC 
experiments observe signatures of unexpected scalar fields.

When we consider  a coordinate transformation that sends
\begin{equation}
\delta \vartheta^m =  - K_L^{am} \varepsilon^a (x^\mu) -  R^m \varepsilon(x^\mu)
\lbl{repna1}
\end{equation}
where $\epsilon^a(x^\mu), \epsilon(x^\mu)$ are arbitrary functions on $\mathbb M^4$,
in direct generalization of (\ref{repa}) we find that the metric (\ref{g7}) 
remains intact provided 
\begin{equation}
\begin{matrix} \delta A_\mu^a  &= & \partial_\mu \varepsilon^a + \epsilon^{abc} A_\mu^b \varepsilon^c \\
\delta B_\mu  & = &  \partial_\mu \varepsilon \ \ \ \ \ \  \ \ \ \ \ \ \ \ 
\end{matrix}
\lbl{repna2}
\end{equation}
This is the  $SU_L(2) \times U_R(1)$ gauge transformation law of the gauge fields $(A^a_\mu, B_\mu)$.

In order to perform the projection to massless states we assume  that 
the $\mathbb S^3_\xi$ metric components $g^\xi_{mn}$ and the components $(K_L^{am}, R^m)$ of the Killing vectors 
depend solely on the three internal coordinates $\vartheta^m$ with no $x^\mu$ dependence, 
while $g_{\mu\nu}$ and $(A^a_\mu, B_\mu)$ all depend only on the four dimensional $x_\mu$.  We 
substitute the metric (\ref{g7}) in (\ref{E7}) and we integrate over $\mathbb S^3_\xi$  to get
\begin{equation}
S = \frac{1}{\mathfrak h} \int d^4x \sqrt{g}\left[ \frac{1}{r^2} \left\{ R +  R_{int}  \right\}  +  \frac{1}{4} \frac{\xi^2 + 2}{3}\vec G_{\mu\nu}\cdot 
\vec G^{\mu\nu} +  \frac{1}{4} \xi^2 F_{\mu\nu}^2  \right]
\lbl{7dE}
\end{equation}
Here $\mathfrak h$ is an {\it a priori} arbitrary  dimensionless number, obtained by combining the various 
overall factors into a single quantity (we may call it a "Planck's constant"). All the metric structure is 
determined by the four dimensional $g_{\mu\nu}$,  and
${\vec G}_{\mu\nu}$ is the $SU(2)$ field strength of ${\vec A}_\mu$ and $F_{\mu\nu}$ is the $U(1)$ field strength
of $B_\mu$. The internal scalar curvature is
\[
R_{int} = \frac{4-\xi^2 }{2r^2}
\]
and it has the r\^ole of  a cosmological constant.

With  (\ref{7dE}), we now wish to recover the  Weinberg-Salam Lagrangian (\ref{egw1}),
(\ref{ewg2}). For this we introduce the locally conformally flat metric tensor with components (\ref{grho})
and substitute in  (\ref{7dE}). The result is
\[
 S \ = \   \frac{1}{\mathfrak h} \int d^4x  \left\{  \left[ \frac{6}{\kappa^2} \frac{1}{r^2} (\partial_\mu \rho)^2 +  \frac{\rho^4}{2r^4 \kappa^4}  (4-\xi^2) 
\right]  +  \frac{1}{4} \frac{\xi^2 + 2}{3} (\vec G_{\mu\nu})^2  + \frac{1}{4} \xi^2 F_{\mu\nu}^2  \right\}
\]
This reproduces the first four terms in (\ref{wsgi})  (up to the overall dimensionless 
factor $\mathfrak h$) when we choose
 the (constant) radius $r^2$ to be
\begin{equation}
r^2 =  \frac{6}{\kappa^2}
\label{parar}
\end{equation}
and we scale the
gauge fields as follows,
\begin{equation}
\begin{matrix}
A^a_\mu \to q A^a_\mu 
\\
B_\mu \to  q' B_\mu \end{matrix}
\lbl{qq}
\end{equation}
where we select
\begin{equation}
q = \sqrt{ \frac{3}{\xi^2 + 2 }}
\label{coupq}
\end{equation} 
\begin{equation}
q' = \frac{1}{\xi}
\label{coupqq}
\end{equation} 
and
\begin{equation}
\lambda =   \frac{1}{4!}\frac{4-\xi^2 }{3 }
\label{parah}
\end{equation}
In particular, these definitions ensure that the Yang-Mills contribution to the action acquires the correct canonical normalization (\ref{ewg2}),
\begin{equation}
S_{YM} = \int d^4x \sqrt{g} \left\{ \frac{1}{4} \vec G_{\mu\nu} \cdot \vec G^{ \mu\nu }  + \frac{1}{4} F_{\mu\nu} F^{\mu\nu}
\right\}
\lbl{normYM}
\end{equation}
Moreover, we note that the $SU_L(2)$ coupling $q$, the $U_R(1)$ coupling $q'$ 
and the Higgs coupling $\lambda$ are now {\it all} determined by the dimensionless squashing parameter $\xi$.

\section{Vector Boson Mass And Nambu Brane}

\vskip 0.5cm

We proceed to construct the gauge invariant mass terms for the intermediate vector bosons. Following Section \ref{vmfb} we  shall
here show how
a mass term can be obtained from  a three-brane with Nambu action.  From a geometrical point of view the Nambu action is
a very natural choice. However, we shall find that it does not conform with the experimentally 
observed $W_\mu^\pm$ and $Z_\mu$ masses.  The reason is that the Nambu action breaks an underlying local  $SU_L(2) \times SU_R(2)$ custodial 
symmetry of the mass matrix. In the next Section we show
how the custodial symmetry is recovered and the correct intermediate vector boson masses obtained.

As in Section \ref{vmfb}  we introduce
a three-brane that stretches along the non-compact directions 
of $\mathbb M^4 \times \mathbb S_\xi^3$.  Locally the brane  is described by
\begin{equation}
\vartheta^m = X^m(x_\mu)
\lbl{htheta2}
\end{equation}
In analogy with (\ref{5bein}) we
introduce the basis vectors on the brane, 
\[
{E_\mu}^m = {\delta_\mu}^m + \frac{\partial X^m }{\partial x^\mu} 
\]
Together with (\ref{g7}) this leads to the induced brane metric
\begin{equation}
G^{ind}_{\mu\nu} = {E_\mu}^\alpha {E_\nu}^\beta g_{\alpha\beta} = 
g_{\mu\nu} + \frac{r^2}{4}  g^\xi_{mn} (K_L^{am} A^a_\mu  + R^m B_\mu + \partial_\mu X^m) (K_L^{bn} A^b_\nu + R^n B_\nu + \partial_\nu X^n)
\lbl{indbrane2}
\end{equation}
in direct generalization of (\ref{indbrane}).
We compute its determinant and the result is
\[
\det [G^{ind}_{\mu\nu}] = 
\]
\begin{equation}
\det [ g_{\mu\nu} ] \cdot \left( 1 + \frac{r^2}{4} g^{\mu\nu}g^\xi_{mn} (K_L^{am} A^a_\mu 
+ R^m B_\mu + \partial_\mu X^m) 
(K_L^{bn} A^b_\nu + R^n B_\nu + \partial_\nu X^n)\right)
\lbl{7detG}
\end{equation}
Here the three composites
\begin{equation}
\mathcal J^m_\mu = {K_L^{am}}A^a_\mu + R^m B_\mu + \frac{ \partial X^m }{\partial x^\mu}
\lbl{JAX}
\end{equation}
are  the brane  versions of the gauge invariant supercurrents
(\ref{J+}), (\ref{J3}). By comparing (\ref{7detG}) with (\ref{g7}) we conclude that these 
supercurrents are indeed invariant under the 
reparametri\-za\-tions (\ref{repna1}), (\ref{repna2}) {\it a.k.a.} $SU_L(2)\times U_R(1)$ gauge 
transformations (recall that together $(K_L^{am} , R^m)$ 
generate the unbroken $SU_L(2)\times U_R(1)$ isometry of $\mathbb S^3_\xi$).
For example, in order to explicitly verify the invariance under the non-Abelian reparametrization (\ref{repna2}) 
we first observe that
\[
\delta (K_L^{am} A^a_\mu) =(  \mathcal L_{-\varepsilon^b K_L^b }K_L^{am})  A_\mu^a
+ {K_L}^{am}(  \partial_\mu \varepsilon^a + \epsilon^{abc}  A^b_\mu \varepsilon^c )
\]
\[
=  \epsilon^{abc}  A^a_\mu  \varepsilon^b K_L^{cm} 
+ {K_L}^{am} \partial_\mu \varepsilon^a + \epsilon^{abc} \varepsilon^a  A^b_\mu 
{K_L}^{cm}  = {K_L}^{am} \partial_\mu \varepsilon^a
\]
On the other hand, from (\ref{htheta2}) we get by (\ref{repna1}) that
\[
\delta \left( \frac{\partial X^m}{\partial x_\mu} \right) = - {K_L}^{am} 
\partial_\mu \varepsilon^a
\]
Furthermore, in line with (\ref{pur})  the last term in (\ref{JAX}) is a pure gauge contribution. For this we
recall (\ref{mc2}) and (\ref{e1e2n}), (\ref{KR}) to find
\[
\mathcal J^m_\mu   {L^a_m} \frac{\tau^a}{2i} \ = \   ( A^a_\mu + B_\mu \, {e_3}^a)  \frac{\tau^a}{2i}  
\ + \ \mathcal U^{-1} \partial_\mu \mathcal U
\]

\vskip 0.3cm
\noindent
In the limit of small brane fluctuations the Nambu action for the brane can be expanded in derivatives of fluctuations and to leading nontrivial
order we get
\begin{equation}
S_{brane}  \ = \ \frac{1}{\mathfrak h} \, T \int d^4x \sqrt{G^{ind}} 
\ \approx \
\frac{1}{\mathfrak h}\,  T \int d^4x \sqrt{ g} \cdot  \left( 1 + \frac{1}{2} \frac{r^2}{4} g^{\mu\nu} 
g^\xi_{mn} \mathcal J^m_\mu \mathcal J^n_\nu + ... \right)
\lbl{braneact2}
\end{equation}
Here the first term contributes to the four dimensional cosmological constant and the second  is the
mass term for the supercurrents. We use (\ref{gGG}) to write the mass term in (\ref{braneact2}) as follows,
\begin{equation}
\frac{T}{8} r^2 g^{\mu\nu} g^\xi_{mn} \mathcal J^m_\mu \mathcal J^n_\nu =  g^{\mu\nu} \left( \frac{T}{8} r^2
E_m^i \delta_{ij} E_n^j \right)  \mathcal J^m_\mu \mathcal J^n_\nu \ = \ g^{\mu\nu} M_{mn} (\xi) \mathcal J^m_\mu \mathcal J^n_\nu
\label{mass1}
\end{equation}
where the $E_m^i$ are the squashed dreibeins (\ref{rhmca})-(\ref{rhmcc}).

Since the mass term involves only  three supercurrents,
one linear combination of the  four gauge fields $A^a_\mu, B_\mu$ remains massless.  To identify the massive and
massless combinations we recall that the Kaluza-Klein reparameterizations {\it a.k.a.}
gauge transformations act transitively and consequently we can (locally) introduce a coordinate transformation
that makes the brane coordinates constants: 
$$
\theta \ =  \  \psi_+ \ = \  \psi_- \  =  \ 0
$$
This amounts  to rotating 
\[
\vec e_1 \ \to \ \left( \begin{matrix} 1 \\ 0 \\ 0 \end{matrix} \right) \ \ \ \ \& 
 \ \ \ \vec e_2 
\ \to \ \left( \begin{matrix} 0 \\ 1 \\ 0 \end{matrix} \right) \ \ \ \& \ \  \ \ \vec e_3 \ \to \ \left( \begin{matrix} 0 \\ 0 \\ 1 \end{matrix} \right)
\]
in  (\ref{e1e2n});
From the point of view of the original Weinberg-Salam model this corresponds to selecting the Unitary Gauge
that always exists locally.  We use (\ref{gGG})  with the explicit realizations  
(\ref{lhmca})-(\ref{lhmcc}) and  the rescaled fields (\ref{qq}) and  diagonalize the mass matrix $M_{mn}(\xi)$
to conclude  that the massless combination is 
\begin{equation}
\mathcal A_\mu \ = \ \frac{ q B_\mu - q' A^3_\mu } {\sqrt{ q^2 + {q'}^2}} = -\sin \theta_W \cdot A_\mu^3 + \cos \theta_W \cdot  B_\mu
\lbl{msl}
\end{equation}
and the massive combinations are
\begin{eqnarray}
W_\mu^+ & = & A_\mu^1 + i A_\mu^2
\lbl{W}
\\
Z_\mu & = & \frac{ q ' B_\mu + q A_\mu^3 }{\sqrt{ q^2 + {q'}^2}} \ =  \cos \theta_W  \cdot A_\mu + \sin \theta_W \cdot B_\mu
\lbl{Z}
\end{eqnarray}
where 
\[
\sin^2 \theta_W = \frac{\xi^2 + 2}{4\, \xi^2 + 2}
\]
so that 
\[
\frac{1}{4} \leq \sin^2 \theta_W \leq 1
\]
and we get from (\ref{braneact2}) the mass term 
\begin{equation}
S_{mass}  \ = \ 
\frac{1}{\mathfrak h} \frac{r^2T}{8 }  \int d^4x \sqrt{ g} \cdot  \left\{ q^2 W^+_\mu W^{\mu -} + \xi^2 (q^2 + {q'}^2) Z_\mu Z^\mu \right\}
\lbl{unitmass}
\end{equation}
By  combining this with (\ref{7dE}), (\ref{normYM})  we get for the entire action  in terms of the rescaled, canonical fields 
\[
S = \frac{1}{\mathfrak h} \int d^4x \sqrt{g} \left\{ \frac{\kappa^2}{6} \left[ R + \left( \frac{6T}{\kappa^2} + \frac{2}{4!} (4-\xi^2) \kappa^2 \right)  \right]
+ \frac{1}{4} \vec G_{\mu\nu}\cdot \vec G^{\mu\nu} + \frac{1}{4}  F_{\mu\nu}F^{\mu\nu} \right.
\]
\begin{equation}
\left. + \frac{3 T}{4\kappa^2} \left\{ q^2 W^+_\mu W^{\mu - } + \xi^2 (q^2 + {q'}^2 ) Z_\mu Z^\mu  \right\} \right\}
\lbl{7dEb}
\end{equation}
When we select the locally conformally flat metric tensor (\ref{grho}) and choose the parameters as in 
(\ref{parar})-(\ref{parah}) and
\[
T = \frac{4}{3} \kappa^4 
\]
we  get a Lagrangian which is very similar in form to the Weinberg-Salam  Lagrangian (\ref{wsgi}), with the 
Higgs coupling 
\[
\lambda = \frac{1}{4!} \frac{100- \xi^2}{3} 
\]
In particular,  in addition of the overall $\mathfrak h$ there are  now only {\it two} independent parameters, 
$\kappa$ that determines the mass scale and
$\xi$ that determines the three couplings $q, q'$ and $\lambda$. The apparent difference between (\ref{wsgi}) and (\ref{7dEb}) is in the 
mass relations, they have the same form only when $\xi =1 $.  But in this case we obtain the experimentally quite distant value 
\begin{equation}
\sin^2 \theta_W \ = \ \frac{1}{2}
\label{wrong}
\end{equation}
for the Weinberg angle.  Since we do not understand how to reconcile these differences we propose  
that the Nambu action is not the one realized in Nature to provide masses for the
intermediate vector bosons.

\section{Custodial Symmetry}

\vskip 0.5cm

The mass matrix (\ref{mass1}) is obtained from the induced metric using the Nambu action, 
and as such it has a very natural geometric origin.
For a {\it generic} value of $\xi$ it also shares the local $SU_L(2) \times U_R(1)$ isometry of 
the squashed three-sphere. But  when  $\xi = 1$ so that the metric tensor coincides with the bi-invariant 
(\ref{nosqu}), the symmetry of the mass matrix  (\ref{mass1}) becomes  extended to the local
$SU_L(2) \times SU_R(2)$ invariance and it can be presented entirely 
in terms of the $\mathbb S^3$ Killing vectors as follows, 
\begin{equation}
M_{mn} \ = \ \frac{T}{8} r^2 L^i_m \delta_{ij} L^j_n \ = \  \frac{T}{8} r^2 R^i_m \delta_{ij} R^j_n \  =  \  \frac{T}{16} 
r^2 ( L^i_m \delta_{ij} L^j_n + R^i_m \delta_{ij} R^j_n) 
\label{Mmn}
\end{equation}
We call this local $SU_L(2)\times SU_R(2)$ symmetry of the mass matrix  (\ref{Mmn}) the {\it custodial symmetry}. 
An unbroken custodial symmetry implies the following familiar relation between the intermediate vector boson masses and the 
Weinberg angle,
\begin{equation}
\sin^2 \theta_W \ = \  \frac{{q'}^2}{q^2 + {q'}^2} \ = \   \frac{\xi^2 + 2}{4\xi^2 + 2}  \ = \ 1 - \frac{M_W^2}{M_Z^2} 
\label{theW}
\end{equation}
We also note that the custodial symmetry can be used to justify  {\it a posteriori} 
the relative normalization of the Killing vectors that we have introduced in 
(\ref{g7}). 

Since the squashed metric tensor (\ref{Rdrei}) can  be represented in terms of the $\mathbb S^3$ Killing vectors
independently of $\xi$ 
we may as well adopt the point of view that since the Killing vectors determine the metric tensor they  are 
more "primitive" and the mass matrix (\ref{Mmn}) is the most natural one also in the case of a squashed three-sphere,
{\it irrespectively}
of the value of $\xi$.

The most general 
mass matrix that breaks the custodial symmetry explicitely while retaining the $SU_L(2) \times U_R(1)$ symmetry is
\begin{equation}
M_{mn}(\eta)  \ = \  \frac{T}{8} r^2 \left( R^1_m R^1_n + R^2_m R^2_n + \eta^2 R^3_m R^3_n \right)
\label{cust1}
\end{equation}
Here $\eta$ is a new parameter which is independent of the squashing parameter $\xi$. 
For  $\eta = 1$ we have the  custodial symmetry that 
 becomes explicitely broken into $SU_L(2) \times U_R(1)$ for $\eta \not=1$. Using the mass matrix (\ref{cust1}) we
 introduce the following (Polyakov-like) brane action
 \[
 S_{brane} \ = \ \frac{1}{\mathfrak h}  T \int d^4 x \sqrt{g} \,  g^{\mu\nu} M_{mn}(\eta) \mathcal J^m_\mu \mathcal J^n_\nu 
 \]
With this  we find instead of (\ref{7dEb})
\[
S = \frac{1}{\mathfrak h} \int d^4 x \sqrt{g} \left\{ 
\frac{\kappa^2}{6}  \left[ R + \frac{2}{4!} (4-\xi^2) \kappa^2 )  \right] +  \frac{1}{4} \vec G_{\mu\nu} \cdot \vec G^{\mu\nu} + \frac{1}{4}  F_{\mu\nu}F^{\mu\nu} 
\right.
\]
\begin{equation}
\left. + \kappa^2 \left[ q^2 W^+_\mu W^{\mu - } + \eta^2 (q^2 + {q'}^2 ) Z_\mu Z^\mu  \right]  \right\} 
\label{Scust1}
\end{equation}
where we continue to exclude a bare cosmological constant. This  Lagrangian gives us the mass relation \cite{ross}
\[
\eta^2 \cos^2 \theta_W \ = \ \eta^2  \frac{{q}^2}{q^2 + {q'}^2} \ = \ \frac{M_W^2}{M_Z^2} 
\]
with  the experimental value \cite{pdg}
\[
\eta^2 \ = \  1.01023 \pm 0.00022
\]
We recall  \cite{pdg} that in the Standard Model the difference to the custodial symmetry value $\eta = 1$ is due to bosonic loops.

We now proceed to inspect the (classical) value of the Weinberg angle (\ref{theW}).  For this we shall approximate
$\eta = 1$. From  (\ref{coupq}) and  (\ref{theW}) we find  in the Thomson limit  the value
\[
q \sin\theta_W \ = \  \sqrt{ \frac{3}{4\xi^2 + 2} }  \ = \ e  \  \ \Rightarrow \  \ \xi  \  = \ \frac{1}{2} \sqrt{\frac{3}{e^2} - 2} \ \approx \ 2.77
\]
so that
\[
\sin^2 \theta_W \ \approx \ 0.296
\]
and from (\ref{coupq}), (\ref{coupqq}) we get
\[
q \ = \ 0.557 \ \ \ \  \& \ \ \ \  q' \ = \ 0.361
\]
These numbers are surprisingly close to the experimental low momentum transfer values (\ref{lowm}), (\ref{expg}) in particular 
when we take into account that the present estimations are purely classical and in particular
we have not taken into account any interactions nor
any fermionic effects.

Furthermore, in the absence of a bare seven-dimensional cosmological constant we get from (\ref{parah}) the numerical
value
\[
\lambda = \frac{1}{4!} \frac{4 - \xi^2}{3} \ = \ - 0.0511
\]
which is  small, but negative; Adding a small but positive bare cosmological constant would make the effective
Higgs coupling positive but  here we prefer to avoid this.  Instead, we note that in the pure scalar $\lambda \phi^4$ 
field theory the four dimensional triviality is well established for bare $\lambda < 0$ \cite{froh} and this suggests that
quantum effects could also here drive $\xi \to 2$.

Suppose now that we are in a conformally flat and Euclidean-Lorentz {\it i.e.} SO(4) invariant classical ground state of
(\ref{Scust1}). The vector fields must all then vanish and when we substitute (\ref{grho}) in (\ref{Scust1}) we obtain the
following equation for the conformal scale of  the metric  tensor (\ref{grho}),
\[
- \Box \left( \frac{\rho}{\kappa} \right) + \frac{\kappa^2}{3 \cdot 4!}  (4 - \xi^2)  \left( \frac{\rho}{\kappa} \right)^3 \ = \ 0
\]
This is solved by
\begin{equation}
ds^2 =  \left( \frac{\rho}{\kappa} \right)^2 \eta_{\mu\nu} dx^\mu dx^\nu \ = \  \frac{ \eta_{\mu\nu} dx^\mu dx^\nu }
{\left[ 1 + \frac{4-\xi^2}{(4!)^2} \kappa^2 \cdot x^2 \right]^2 }
\label{DFFm}
\end{equation}
This gives us either the de Sitter or anti de Sitter metric as the ground state, depending on whether $\xi < 2$ or $\xi > 2$. These could
be viewed as two different phases of the theory,  and the tricritical  value
$\xi = 2$ yields a flat $\mathbb R^4$ and corresponds to a Weinberg angle value
\[
\sin^2 \theta_W \ = \ \frac{1}{3}
\]
We note that  according to our model this means that the Grand Unified prediction for the Weinberg angle  \cite{dine}
\[
\sin^2 \theta_W \ = \ \frac{3}{8}
\]
corresponds to  a phase which is {\it different} from the observed world.

Finally, since the (anti) de Sitter manifold is homogeneous and has constant curvature, we obtain a reasonable ground state
expectation value for $\rho$ assuming that we are in the vicinity of $x = 0$  in (\ref{DFFm}). This  yields the estimate
\[
<\rho> \ \approx \ \kappa
\]
and gives us the standard relation between the value of the Higgs condensate and the intermediate vector boson masses.
We conclude by noting, that in the quantum theory there will be corrections to this expectation value due to the intermediate
vector boson condensates (\ref{Zakh1}), (\ref{Zakh2}).

\section{The limit of Two-Sphere  }
\label{lasts}

\vskip 0.5cm

In this Section we  briefly consider the limit $\xi \to 0$ in the metric (\ref{squ}). This is of interest, since the limit  represents a submanifold of the
squashed three sphere that is the smallest manifold that allows a realization of the $SU(2) \times U(1)$ Lie algebra in terms of Killing vectors.
In this limit we obtain the standard metric of 
$\mathbb S^2 \in \mathbb R^3$ 
\[
ds^2 =  \frac{r^2}{4} g_{mn} d\vartheta^m d\vartheta^n =
\frac{r^2}{4} \left\{ (d\theta)^2 + \sin^2 \theta (d\psi_-)^2 \right\}
\]
The dreibein (\ref{lhmca})-(\ref{lhmcc}) becomes
\begin{eqnarray*}
L^1  & \to  & {e_3}^1 \cos \theta d\psi_-  -  {e_2}^1 d\theta 
\\
L^2 & \to &  {e_3}^2 \cos\theta d\psi_-  -  {e_2}^2 d\theta  
\\
L^3 & \to  & {e_3}^3 \cos\theta d\psi_- - d\psi_- \
\end{eqnarray*}
These are the dual to the three dimensional angular momentum operators $l^a$ in (\ref{KT1})-(\ref{KT3})
with respect to the $\mathbb S^2$ metric,
\[
L^a_m = g_{mn} l^{an}
\]
We remind that the two sphere is the coadjoint orbit of $SU(2)$ and so it supports
a representation of   $SU(2)$ which is given by the $\mathbb S^2$ 
Killing vectors {\it a.k.a.}  angular momentum operators $l^a$.  

When we send $\xi \to 0$ in the Lagrangian (\ref{7dEb}) and remove the $\psi_+$ dependence, again
rotating $\vec e_3$ to point towards the north pole 
we get
\begin{equation}
S = \frac{1}{\mathfrak h}  \int d^4x \sqrt{-g} \left\{ \frac{1}{r^2}\left[ R + R_{int}  \right]
+ \frac{1}{4} \vec G_{\mu\nu}\cdot \vec G^{\mu\nu} + \frac{24\pi G T}{8} W_\mu W^{\mu \star} 
\right\}
\lbl{7d2sphe}
\end{equation}
We observe that  only two components of the $SU(2)$ gauge field are massive. This is the result we expect to have
when we break $SU(2)$ into $U(1)$ in an Non-Abelian Higgs model, with the Higgs field in the adjoint representation
of $SU(2)$. 

\vskip 1.0cm
\section{Conclusions}
 
 \vskip 0.5cm
 
We have addressed the electroweak hierarchy problem posed by the instability of the Higgs 
mass by inspecting whether an apparent general covariance but with a locally conformally flat metric tensor 
that is present in the  Weinberg-Salam Lagrangian, could  be somehow 
interpreted in terms of  full general covariance in  a higher  dimensional gravity theory. We have argued  that if one  starts from a seven 
dimensional generally covariant 
action that has the same form as the Hilbert-Einstein action and projects on a subset of its Kaluza-Klein decomposed fields,  one arrives at the functional form  of the 
Weinberg-Salam Lagrangian with correct vector boson masses but with no bare Higgs mass,  as the presence
of the Higgs mass is forbidden by dimensional censorship as it breaks general covariance of the gravity theory.  Moreover, if the quantization
of the electroweak theory can be performed so that a Ward-Takahashi like identity that ensures the preservation
of the higher dimensional general covariance, the Higgs mass remains absent and that could provide a resolution
to the electroweak hierarchy problem. Indeed, it has been argued that the absence of a bare Higgs mass
could help to resolve the gauge hierarchy problem \cite{man} and it has also been
argued that despite of the absence of the bare mass the  eventual Higgs expectation value does not need
to be small but can acquire a realistic value \cite{elias}.  We have here shown that at the 
classical level this could be due to the (anti) de Sitter ground state
of the theory, while in the quantum theory a nontrivial expectation value for the Higgs condensate
could emerge from the $A^2$ condensation
of the intermediate vector bosons. 

This line of arguments, if validated in the context of electroweak theory,
would give us the first indication that four dimensional  physics is sensitive to higher dimensional symmetry structure, 
even though it may remain questionable whether the Weinberg-Salam model with its fermions indeed emerges 
from a Kaluza-Klein reduction of some complete and internally consistent  higher dimensional theory theory. 

In addition, we have found that the higher dimensional general covariance enforces us a relation between the two 
electroweak gauge couplings, and the Higgs self-coupling.  They  are all determinants of a single parameter, the 
squashing parameter of the internal three sphere.  Furthermore, when we use the known low energy 
value $1/137$ of the electromagnetic fine structure constant our approach predicts the value $\sin^2\theta_W = 
0.296$ for the Weinberg angle. Potentially this value could be brought even closer to the observed value by 
inclusion of  interactions, quantum effects and fermions, thus it may serve as an experimental test of the validity
of our higher dimensional approach.

An interesting peculiarity in our approach is the absence of
a primordial Higgs field.  The modulus of Higgs field resides in the conformal scale
of the four dimensional metric and as such it has no direct r\^ole in the mass generation of the vector fields.
Instead the intermediate vector bosons  acquire their masses from a three brane that asymptotically coincides
with the physical space-time but is locally allowed to fluctuate into higher dimensions. 

Except for the relation between the coupling constants that should eventually become an experimental
test between our approach and the standard electroweak theory, the phenomenological content of the present Kaluza-Klein 
based electroweak theory appears to be very similar to that of the conventional Weinberg-Salam model.  But we 
also note that there could be subtle differences \cite{feno} that might become visible at the LHC experiments. 
In particular, the potential observation of additional neutral scalar particles at LHC besides the modulus of the 
Higgs could have an interpretation either in terms of  the two  non-conformal modes that describe the physical 
field degrees of freedom of our four dimensional Hilbert-Einstein action (\ref{Scust1}) or in terms of the higher dimensional 
dilaton fields that are present if the  higher dimensional generally covariant  theory is interpreted  literally in the conventional 
Kaluza-Klein sense.

\vskip 1.0cm

\section*{Acknowledgements}

This work has been supported by a grant from VR (Vetenskapsr\.adet), and by a 
STINT Institutional grant IG2004-2 025.  We both thank Maxim Chernodub for discussions. A.J.N. thanks Joe  Minahan and
Sergey   Solodukhin,   and S.S.  thanks Ulf Danielsson,  Susha Parameswaran and  Konstantin Zarembo for discussions.


\vskip 1.0cm

\begin{thebibliography}{99}


\bibitem{wel09} J.D. Wells, {\it Lectures  on Higgs Boson Physics in the Standard Model and Beyond}, E-print
{\tt arXiv:0909.4541v1} [hep-ph]

\bibitem{pol92}  J. Polchinski,  	{\it Effective Field Theory and the Fermi Surface}, E-print {\tt arXiv:hep-th/9210046v2} [hep-th]

\bibitem{giu08}  G. Giudice, {\it Naturally Speaking: The Naturalness Criterion and Physics at the LHC}  E-print {\tt 	arXiv:0801.2562v2} [hep-ph]
 
\bibitem{dine} M.~Dine, {\it Supersymmetry and String Theory: Beyond the Standard Model} (Cambridge University Press, Cambridge, 2007)

\bibitem{little} M.  Schmaltz and D. Tucker-Smith, Ann. Rev. Nucl. Part. Sci. {\bf 55} 229 (2005)

\bibitem{sun} L. Randall and  R. Sundrum,  Phys.  Rev.  Lett. {\bf  83} 3370 (1999)

\bibitem{pope} M.J. Duff, B.E.W. Nilsson and C.N. Pope, Phys. Rept. {\bf C130}, 1 (1986)

\bibitem{appel}  T. Appelquist, A. Chodos and P.G.O. Freund, {\it Modern Kaluza-Klein Theories}  (Addison-Wesley, Menlo Park, 1987)

\bibitem{witten} E. Witten, Nucl. Phys. {\bf B186} 412 (1981)

\bibitem{guba1} F.V. Gubarev and V. I. Zakharov,  Phys. Lett. {\bf B501} 28 (2001)

\bibitem{guba2} F.V. Gubarev, L. Stodolsky and V.I. Zakharov,   Phys. Rev. Lett. {\bf86} 2220 (2001)

\bibitem{fad2}  L.D. Faddeev and A.J. Niemi, Nucl. Phys. {\bf B776}, 38 (2007) 

\bibitem{E158} P.L. Anthony {\it et.al.}, Phys. Rev. Lett {\bf 95}, 081601  (2005)

\bibitem{pdg}  C. Amsler et al. (Particle Data Group), Phys. Lett.  {\bf B667}, 1 (2008) 

\bibitem{gibb} G.W. Gibbons, S.W. Hawking and M.J. Perry, Nucl. Phys. {\bf B138} 141 (1978)

\bibitem{shapo} V.V. Vlasov, V.A. Matveev, 
A.N. Tavkhelidze, S.Y. Khlebnikov and
M.E. Shaposhnikov, Fiz.\ Elem.\ Chast.\ Atom.\ Yadra {\bf 18}  5 (1987)

\bibitem{volo} G.E. Volovik and T. Vachaspati, Int. J.  Mod.  Phys. {\bf B10}  471 (1996)

\bibitem{fad}  M.N. Chernodub, L.D. Faddeev and A.J. Niemi,  JHEP 0812:014,2008;  M.N. Chernodub and  A.J. Niemi,
Phys. Rev. {\bf D77}, 127902 (2008) 

\bibitem{lude} L.D. Faddeev, e-Print: arXiv:0811.3311 [hep-th] 

\bibitem{max}  M.N. Chernodub  and E.-M. Ilgenfritz,  Phys. Rev. {\bf D78} 034036 (2008) 

\bibitem{polch} J. Polchinski, Nucl. Phys. {\bf B303}, 226 (1988)

\bibitem{poly} A. Polyakov,  Int. J. Mod. Phys. {\bf A16}, 4511 (2001)

\bibitem{sund} R. Sundrum, e-Print: hep-th/0312212

\bibitem{foot1} We recall the familiar result  that in a quantum theory special conformal symmetry commonly becomes broken by Weyl anomaly, with  the trace of
the energy momentum tensor acquiring  a non-vanishing expectation value \cite{foot2}. For example in Einstein-Yang-Mills theory
\[
<{T^\mu}_\mu> \ \propto \ a Tr \{ F_{\mu\nu}^2 \} + b R^2 + ...
\]
The potential relation between Weyl anomaly and Higgs mass in the present context remains to
be clarified.


\bibitem{foot2} D.M. Capper, M.J. Duff and L. Halpern, Phys. Rev. {\bf D10} 461 (1974)

\bibitem{abers}
E.S. Abers and B.W. Lee, Phys. Rept. {\bf C9} 1 (1973).

\bibitem{cho} Y.M. Cho and P.G .O.  Freund, Phys. Rev. {\bf D12},  1711 (1976)

\bibitem{sergey} S. Slizovskiy, e-Print: arXiv:1004.0216

\bibitem{ross}  D. Ross and M. Veltman,  Nucl. Phys. {\bf B95}, 135 (1975)

\bibitem{froh} R. Fern\'andez, J. Fr\"ohlich and A.D. Sokal, {\it Random Walks, Critical Phenomena and Triviality
in Quantum Field Theory} (Springer-verlag, New York, 1992)

\bibitem{man} L. Alexander-Nunneley and  A. Pilaftsis, e-Print: arXiv:1006.5916 [hep-ph]

\bibitem{elias} F.A. Chishtie, T. Hanif, J. Jia, R.B. Mann, D.G.C. McKeon, T.N. Sherry and T.G. Steele, 
e-Print: arXiv:1006.5887 [hep-ph]

\bibitem{feno} M.G. Ryskin and A.G. Shuvaev,  e-Print: arXiv:0909.3374 [hep-ph] 



\end{thebibliography}
\end{document}